\documentclass[12pt, a4paper]{article}
\pdfoutput=1
\usepackage{cite}
\usepackage{amsmath}
\usepackage{amssymb}
\usepackage{graphicx}
\usepackage{rotating}
\usepackage{bm}
\usepackage{color}
\usepackage{subfigure}
\usepackage{colortbl}
\usepackage{url}
\newcount\timecount
\newcount\hours \newcount\minutes  \newcount\temp \newcount\pmhours
\hours = \time
\divide\hours by 60
\temp = \hours
\multiply\temp by 60
\minutes = \time
\advance\minutes by -\temp
\def\hour{\the\hours}
\def\minute{\ifnum\minutes<10 0\the\minutes
            \else\the\minutes\fi}
\def\clock{
\ifnum\hours=0 12:\minute\ AM
\else\ifnum\hours<12 \hour:\minute\ AM
      \else\ifnum\hours=12 12:\minute\ PM
            \else\ifnum\hours>12
                 \pmhours=\hours
                 \advance\pmhours by -12
                 \the\pmhours:\minute\ PM
                 \fi
            \fi
      \fi
\fi
}

\def\monthname{\relax\ifcase\month 0/\or January\or February\or
   March\or April\or May\or June\or July\or August\or September\or
   October\or November\or December\else\number\month/\fi}

\def\bold#1{\setbox0=\hbox{$#1$}%
     \kern-.025em\copy0\kern-\wd0
     \kern.05em\copy0\kern-\wd0
     \kern-.025em\raise.0433em\box0 }

\definecolor{Orange}{cmyk}{0,0.61,0.87,0}
\definecolor{JungleGreen}{cmyk}{0.99,0,0.52,0}
\definecolor{OliveGreen}{cmyk}{0.64,0,0.95,0.40}
\definecolor{Brown}{cmyk}{0,0.70,1,0.40}
\definecolor{RoyalBlue}{cmyk}{0.71,0.53,0,0.12}
\definecolor{Gray}{cmyk}{0,0,0,0.40}
\definecolor{LightPink}{cmyk}{0.0,0.25,0,0}
\definecolor{LLightPink}{cmyk}{0.0,0.10,0,0}
\definecolor{LightBlue}{cmyk}{0.25,0,0,0}
\definecolor{LightGray}{cmyk}{0,0,0,0.2}

\def\beq{\begin{equation}}
\def\eeq{\end{equation}}
\def\beqn{\begin{eqnarray}}
\def\eeqn{\end{eqnarray}}

\def\ga{\mathrel{\raise.3ex\hbox{$>$\kern-.75em\lower1ex\hbox{$\sim$}}}}
\def\la{\mathrel{\raise.3ex\hbox{$<$\kern-.75em\lower1ex\hbox{$\sim$}}}}
\def\gev{{\rm \, Ge\kern-0.125em V}}
\def\tev{{\rm \, Te\kern-0.125em V}}
\def\gyr{{\rm \, G\kern-0.125em yr}}

\def\gappeq{\mathrel{\rlap {\raise.5ex\hbox{$>$}}
{\lower.5ex\hbox{$\sim$}}}}
\def\lappeq{\mathrel{\rlap{\raise.5ex\hbox{$<$}}
{\lower.5ex\hbox{$\sim$}}}}
\def\Toprel#1\over#2{\mathrel{\mathop{#2}\limits^{#1}}}

\def\m12{m_{1\!/2}}

\def\bea{\begin{eqnarray}}
\def\eea{\end{eqnarray}}

\usepackage{multirow}
\newcommand{\model}[3]{${\tt #1}_{\tt\bf #2}^{\tt #3}$}
\newcommand{\DM}[2]{${\tt #1}_{\tt #2}$}

\begin{document}
\begin{titlepage}
\pagestyle{empty}
\baselineskip=21pt
\begin{flushright}
UT--16--31,
UMN--TH--3612/16, FTPI--MINN--16/32
\end{flushright}
\vskip 0.5in
\begin{center}
{\Large{\bf Asymmetric Dark Matter Models in SO(10)}}
\end{center}
\begin{center}
\vskip 0.3in
{\bf Natsumi Nagata}$^{1}$,
{\bf Keith~A.~Olive}$^{2,3}$
and {\bf Jiaming Zheng}$^{2}$
\vskip 0.3in
{\small {\it
$^1$Department of Physics, University of Tokyo, Bunkyo-ku, Tokyo
 113--0033, Japan
\\
\vspace{0.25cm}
$^2${School of Physics and Astronomy, University of Minnesota,
 Minneapolis, MN 55455, USA}\\ 
  \vspace{0.25cm}
$^3${William I. Fine Theoretical Physics Institute, School of Physics
 and Astronomy,\\ 
  \vspace{-0.25cm}
University of Minnesota, Minneapolis, MN 55455, USA}
}}

\vskip 0.5in
{\bf Abstract}
\end{center}
\baselineskip=18pt \noindent

 We systematically study the possibilities for asymmetric dark matter in
 the context of non-supersymmetric SO(10) models of grand unification.
 Dark matter stability in SO(10) is guaranteed by a remnant
 $\mathbb{Z}_2$ symmetry which is preserved when the intermediate scale
 gauge subgroup of SO(10) is broken by a {\bf 126} dimensional
 representation. The asymmetry in the dark matter states is directly 
 generated through the out-of-equilibrium decay of particles around the
 intermediate scale, or transferred from the baryon/lepton asymmetry
 generated in the Standard Model sector by leptogenesis. We
 systematically classify possible asymmetric dark matter candidates in
 terms of their quantum numbers, and derive the conditions for each case
 that the observed dark matter density is (mostly) explained by the
 asymmetry of dark matter particles.

\vskip 0.2in
\leftline{November 2016}
\end{titlepage}
\baselineskip=18pt

\section{Introduction}
\label{sec:intro}

Using only six parameters, the so-called concordance model of standard
Big Bang cosmology does remarkably well in describing the observed
Universe as seen for example in recent cosmic microwave background (CMB)
\cite{Planck}, supernovae \cite{SN}, and baryon acoustic oscillation
(BAO) \cite{bao} data. Among these parameters are the density of
baryons, $\Omega_B h^2 = 0.022$, and the density of  cold dark
matter, $\Omega_c h^2 = 0.12$, where quoted values are taken from
Ref.~\cite{Planck}. Most studies attempting to explain the density of
dark matter assume thermal initial conditions and compute the relic
density remaining after the freeze-out of primordial annihilations. 
The thermal freeze-out mechanism was first used to set constraints on
(then) possible heavy neutrino masses \cite{hut,lw,vdz}. Updated
calculations \cite{ko,swo} for heavy neutrino dark matter candidates
provided lower bounds of 3--7~GeV depending on whether the neutrinos had
Dirac or Majorana masses and details of the quark hadron transition in
the early Universe \cite{swo,oqh}.

Asymmetric dark matter first emerged as a means for complementing the
above limits on neutrino masses \cite{ho}. For the case of Dirac
neutrino masses, an asymmetry between neutrinos and anti-neutrinos allows
one to place an upper limit on the neutrino mass. If the asymmetry is
of order the baryon asymmetry this upper limit is  approximately 60~GeV,
and it was demonstrated in Ref.~\cite{ho} that the out-of-equilibrium
decay scenario (the leading mechanism for generating a baryon asymmetry
at the time) would indeed provide an asymmetry in a fourth generation
neutrino comparable to the baryon asymmetry. The possible connection
between the relatively similar baryon and dark matter densities has
since been a motivating factor for current models of asymmetric dark
matter \cite{adm, Ibe:2011hq, Blum:2012nf, Servant:2013uwa, Boucenna:2015haa,
Dhen:2015wra} (for reviews of asymmetric dark matter, see
Ref.~\cite{admreviews}).   

While Standard Model (SM)-like neutrinos are no longer viable dark
matter candidates, there are many other possibilities for asymmetric
dark matter. Here, we consider the possibility for asymmetric dark
matter in the context of non-supersymmetric SO(10) models of grand
unification \cite{kkr,kkr2, moqz, mnoqz, Brennan:2015psa, noz,
Arbelaez:2015ila, Boucenna:2015sdg, Evans:2015cqq, mnoz, Parida:2016hln}
(GUTs). In SO(10) models with an intermediate scale broken by a {\bf
126} dimensional representation of SO(10), a discrete $\mathbb{Z}_2$
symmetry is preserved at low energies \cite{Kibble:1982ae}. In this
case, a scalar belonging to either a {\bf 16} or {\bf 144} will be
stable if it is the lightest member of the multiplet. In addition, these
models may allow for the unification of the gauge couplings at the GUT
scale \cite{GN2,so10gc}, with a sufficiently long proton lifetime and
acceptably low neutrino masses \cite{noz}. It has also been shown that
models with scalar singlet dark matter protect the Higgs quartic
coupling from running negative, thus preserving the stability of the
electroweak vacuum, and can trigger radiative electroweak symmetry
breaking at the weak scale \cite{mnoz}. 

In this work, we consider the possibility that the SO(10) dark matter candidate
possesses an asymmetry which accounts for the observed relic density.
The asymmetry in the dark matter states can be produced via the
out-of-equilibrium decay of intermediate-scale particles, or by
transferring a part of the asymmetry in the SM sector, which is
generated by leptogenesis \cite{fy}, to the dark matter
sector. Preservation of this asymmetry in the background of sphaleron
interactions, and possible dark matter-anti matter oscillations will
impose stringent constraints on the possible models.  

We show that in models with a minimal field content below the GUT scale,
only scalar singlet dark matter models can be viable. Other models 
either do not allow for phenomenologically acceptable gauge coupling unification
or are hypercharged and require mass splittings which are incompatible
with other derived constraints. For the scalar singlet case, we show
that an asymmetry produced by a transfer process requires a relatively
low intermediate scale which is not possible in minimal models. In
contrast, by including a fermionic singlet, cogenesis models which
produce the dark matter asymmetry directly through out-of-equilibrium
decay are possible.

We also consider some next-to-minimal models which contain
additional fields. While the dark matter candidate itself is not
asymmetric (it is in fact a Majorana fermion in both cases considered),
its relic density is determined by an asymmetry produced by either
cogenesis or transfer into one of the the additional states which
subsequently decays to the dark matter candidate.

In the ensuing discussion, we will first briefly describe the SO(10)
models under consideration. A generic argument for the generation of
asymmetries in the dark matter, baryon, and lepton numbers in the SO(10)
models is given in Sec.~\ref{sec:dmmodels}. We also discuss in
Sec.~\ref{sec:dmmodels} some generic problems associated with asymmetric
hypercharged dark matter. In section~\ref{sec:scalarsinglet}, we
concentrate on the possibility for asymmetric dark matter with scalar
singlets. Here, we consider asymmetries produced by either thermal
transfer or through cogenesis. To try expand the universe of asymmetric
models, we consider some non-minimal models in
section~\ref{sec:NMmodels}. Our conclusions will be given in
section~\ref{sec:conclusion}.

\section{Asymmetric dark matter in SO(10)}
\label{sec:dmmodels}

\subsection{SO(10) dark matter models}
\label{SO(10)models}

As noted above, with the exception of the hierarchy problem,
non-supersymmetric SO(10) GUT \cite{GN2,so10} models
may contain many of the features often used to motivate supersymmetric
models of dark matter. The SO(10) models we consider here all feature an
intermediate scale gauge subgroup of SO(10), which allows for the
possibility for gauge coupling unification. The specific intermediate
gauge group is determined by the vacuum expectation value (VEV) of the Higgs
representation, $R_1$, breaking SO(10), which may be in a {\bf 45, 54}
or {\bf 210} representation.\footnote{In this paper, we restrict our
attention to SO(10) irreducible representations with dimensions up to
210. } 

We employ a {\bf 126} to break the intermediate gauge group at the
intermediate scale $M_{\rm int}$ down to the SM in order to preserve a
$\mathbb{Z}_2$ symmetry related to matter parity. The coupling of the
{\bf 126} to SM matter fields embedded in a {\bf 16} representation of
SO(10) naturally gives rise to a Majorana mass to the singlet
component of the {\bf 16}, $\nu_R^c$,
 of order $\langle {\bf 126} \rangle \sim
M_{\rm int}$, which when combined with the Dirac mass arising from the
VEV of the SM Higgs (now residing in a {\bf 10}-plet of SO(10)) gives
rise to the seesaw mechanism for light neutrino masses
\cite{seesaw}. Furthermore, the out-of-equilibrium decay of the
right-handed neutrinos may yield a lepton asymmetry (leptogenesis
\cite{fy}), which is transferred into a baryon asymmetry through
electroweak sphaleron effects \cite{Manton:1983nd, Kuzmin:1985mm}.

\begin{table}[t]
 \begin{center}
\caption{\it Partial list of ${SU(2)}_L\otimes{U(1)}_Y$ multiplets in
  SO(10) representations that contain an electric neutral color singlet. }
\label{cand}
\vspace{5pt}
\begin{tabular}{l c c l l}
\hline
\hline
 Model & $B-L$ & ${\rm SU}(2)_L$ & $Y\qquad$ &  SO(10) representations\\
\hline
\model{S}{1}{0} &1
 & {\bf 1} & $0$ &   {\bf 16}, {\bf 144} \\
\model{S}{2}{1/2}& 1 & {\bf 2} & $1/2$&    {\bf 16}, {\bf 144} \\
\model{S}{3}{0}& 1 & {\bf 3} & $0$&    {\bf 144} \\
\model{S}{3}{1}& 1 & {\bf 3} & $1$&    {\bf 144} \\
\hline
\model{F}{2}{1/2} & 0 & {\bf 2} & $1/2$ & {\bf 10}, {\bf 120}, ${\bf
		 126}$ ${\bf 210}^\prime$ \\ 
\model{F}{3}{1} & 0 & {\bf 3} & $1$ & {\bf 54} \\ 
\model{F}{4}{1/2} & 0 & {\bf 4} & $1/2$ & ${\bf 210}^\prime$ \\ 
\model{F}{4}{3/2} & 0 & {\bf 4} & $3/2$ & ${\bf 210}^\prime$ \\ 
\model{\widehat{F}}{1}{0}&2 
 & {\bf 1} & $0$ &   {\bf 126} \\
\model{\widehat{F}}{2}{1/2}& 2& {\bf 2} & $1/2$&   {\bf 210} \\
\model{\widehat{F}}{3}{1}&2 & {\bf 3} & $1$& {\bf 126} \\
\hline
\hline
\end{tabular}
 \end{center}
\end{table}

Possible choices for dark matter candidates were discussed in detail in
Ref.~\cite{noz}. Here we will restrict our attention to the candidates
given in Table~\ref{cand}, namely, those who have a non-zero $B-L$ or
hypercharge. Here, each dark matter multiplet is specified by its
spin (${\tt S}$ for scalars and ${\tt F}$ for fermions), SU(2)$_L$
representation ${\bf n}$, and hypercharge $Y$, where the $Q=T_3+Y = 0$
component is the dark matter candidate. 
In general, a fermionic dark matter candidate
should be parity even and belong to a ${\bf 10}$, ${\bf 45}$, ${\bf 54}$,
${\bf 120}$, ${\bf 126}$, ${\bf 210}$ or ${\bf 210}^{\prime}$
representation, while scalar dark matter is parity odd and belongs to a
$\bf 16$ or $\bf 144$ representation. Depending on the dark matter and
Higgs representation chosen, renormalization group evolution of the
gauge couplings can be used to determine the GUT scale, the intermediate
scale, and the value of the GUT gauge coupling. 
We also note that the presence of the intermediate gauge symmetry, as
well as dark matter fields which couple to the SM Higgs field, may
prevent the Higgs quartic coupling from running negative at high energies.
In fact, it has been
shown \cite{mnoz} that for models in which the dark matter is a scalar
singlet, the stability of the Higgs vacuum can be ensured and the
electroweak symmetry may be broken radiatively as in supersymmetric
models.

\subsection{Generation of asymmetries}
\label{sec:genofasym}

In the SO(10) GUT models we consider in this paper, the $B-L$ symmetry
is spontaneously broken at the intermediate scale $M_{\rm int}$ by the
VEV of a {\bf 126} field. At this scale, some of the components of a
representation of the intermediate gauge symmetry acquire masses of
${\cal O} (M_{\rm int})$. For example, the singlet component of the {\bf
126} field, which sits inside a $(\overline{\bf 10}, {\bf 1}, {\bf 3})$
if the intermediate gauge symmetry is ${\rm SU}(4)_C \otimes {\rm
SU}(2)_L \otimes {\rm SU}(2)_R$, obtains a mass from its own VEV, just
like the SM Higgs boson having a mass proportional to its VEV. The
right-handed neutrinos are also such particles as they obtain masses
from the {\bf 126} VEV through their Yukawa couplings. 

The decay of such intermediate-scale particles can generate a $B-L$
asymmetry. The $B-L$ charge in the decay process may not be conserved if
the relevant diagrams contain the {\bf 126} VEV. C and CP
invariance can also be violated if the vertices in the diagrams include CP
phases. Thus, if this decay occurs out-of-equilibrium, a $B-L$ asymmetry
can be generated. A well-known example is the generation of a
lepton-number (and thus $B-L$) asymmetry via the out-of-equilibrium
decay of right-handed neutrinos---leptogenesis \cite{fy,luty}. If there
are no other $B - L$ violating processes in equilibrium, the generated
$B-L$ asymmetry remains non-zero, which results in non-zero baryon and
lepton asymmetries with the help of electroweak sphaleron processes
\cite{Manton:1983nd, Kuzmin:1985mm}.

An asymmetry in the dark matter sector can also be generated through the
production of the $B-L$ asymmetry. We can divide such generation
mechanisms into two types. First, if
intermediate-scale particles can decay not only into SM particles
but also dark matter particles, their out-of-equilibrium decay can directly
produce a $B-L$ asymmetry in the dark matter sector. If this $B-L$
asymmetry is not communicated to the SM sector in the thermal
equilibrium, it remains in the dark matter sector independently, which
results in an asymmetry in the dark matter 
number. The simultaneous generation of asymmetries in the dark
matter, baryon, and lepton numbers is in fact considered in the original
work \cite{ho}, where heavy neutrino dark matter candidates were shown
to have an asymmetry similar to the baryon asymmetry. 
More recently, this scenario has been dubbed cogenesis and
has been discussed widely in the literature \cite{Falkowski:2011xh}.

Second, even if the dark matter sector did not obtain an asymmetry at
the outset, the
$B-L$ asymmetry in the SM sector, which is generated via leptogenesis,
may be transferred to the dark matter sector. For this to occur, some
interactions that communicate the asymmetries between these
sectors should be in thermal equilibrium after the leptogenesis. As we
shall see below, in this case the thermalization conditions give strong
constraints on dark matter models.

In the subsequent sections, we discuss each of these two
scenarios.\footnote{
Actually, there is another type of scenario; an asymmetry is only
produced in the dark matter sector, not in the SM sector, and the
dark matter asymmetry is transmitted to the SM sector via some
interactions so that the observed baryon asymmetry is realized
\cite{Shelton:2010ta}. In SO(10) GUTs, however, leptogenesis is
formulated quite naturally (see, for instance,
Ref.~\cite{Gherghetta:1993kn}), and thus we expect some $B-L$ asymmetry
in the SM sector. For this reason, we do not argue this case in this
paper.  
} The first case is highly dependent on models as
it relies on non-thermal processes at high energies. In the second
case, on the other hand, there are some model-independent aspects, which
we now discuss, that must be shared by all of our SO(10) dark matter
candidates.  

In the transfer scenario, some interactions which distribute $B-L$ between both
the dark matter and SM sectors are required to be in thermal
equilibrium. To derive constraints from this requirement, we first
obtain the temperature associated with leptogenesis.

The tree-level decay rate for a heavy neutrino state $N_i$ (which are
mainly right-handed neutrinos), with the Yukawa coupling matrix, $y$, to
its left-handed counterpart and the Higgs boson is given by 
\begin{align}
 \Gamma_{D_i} = \Gamma (N_i \to H + \ell) +\Gamma (N_i \to H^* +
 \bar{\ell}) = \frac{1}{8\pi} \left(yy^\dagger\right)_{ii} M_i ~,
\end{align}
where $M_i$ are the masses of the heavy neutrino states.

If right-handed neutrinos are thermally produced in the early Universe
and their decay occurs out-of-equilibrium, a maximal lepton asymmetry of
order  
\begin{equation}
 \frac{n_L}{s} \sim \frac{\epsilon}{g_*}
\end{equation}
is generated, where $n_L$ is the net lepton density, $s$ is the entropy
density, $g_* \simeq 100$ is the number of degrees of freedom in the
thermal plasma,  
and $\epsilon$ is a measure of the C and CP violation in the
decay, which is given by
\begin{equation}
 \epsilon \simeq \frac{3}{16\pi}\frac{1}{\left(yy^\dagger\right)_{11}}
 \sum_{i=2,3} {\rm Im} \left[\left(yy^\dagger\right)^2_{i1}\right] 
\frac{M_1}{M_i} ~,
\end{equation} 
for $M_1 \ll M_2, M_3$. 
Note that a non-zero $\epsilon$ is obtained only if we include
loop decay processes with at least two generations
of massive right-handed neutrinos. 
If, on the other hand, the right-handed neutrinos are produced at the end of inflation, the
asymmetry can be related directly to the reheat temperature, $T_R$, by
\cite{dlnos} 
\beq
\frac{n_L}{s}  \sim \epsilon \frac{n_{\nu_R}}{T_R^3} \sim \epsilon f \frac{n_\eta}{T_R^3}
\sim \epsilon f \frac{T_R}{m_\eta} \, ,
\eeq
where $n_\eta$ is the number density of inflatons at the time of their decay,
$f$ is the branching fraction into $\nu_R$, and $m_\eta$ is the inflaton
mass. The produced lepton (or $B-L$) asymmetry then generates baryon
asymmetry as we see in Sec.~\ref{seq:thmlcond}. 

In both of the cases, the produced right-handed neutrinos should decay
out-of-equilibrium. 
In general, the condition for the out-of-equilibrium decay of $N_i$ is
\begin{equation}
 \alpha_{y_i} M_P \lesssim  \mathcal{C} M_i ~,
\label{decay}
\end{equation}
where $\alpha_{y_i} \equiv \left(y y^\dagger\right)_{ii}/8\pi$, $M_P =
1.2 \times 10^{19}$~GeV is the Planck mass, and $\mathcal{C} \equiv (8
\pi^3 g_*/90)^{1/2} \simeq 16.6$ assuming that the number of degrees of
freedom is $g_* \simeq 100$.  
Clearly the time of decay, and hence the time of the generation of the
lepton asymmetry, will be determined by the condition \eqref{decay}. 
Let us define the parameter, $\xi \equiv \alpha_{y_1} M_P/M_1$.
One can show that there is a critical value for $\xi$ such that the
right-handed neutrinos decay in either a matter or radiation dominated
expansion 
\begin{equation}
\begin{cases}
\xi \ll \xi_c & \text{Matter domination} \\[2pt]
\xi \gg \xi_c & \text{Radiation domination} 
\end{cases}
\, ,
\end{equation}
where 
\beq
\xi_c \equiv \frac{2^{9/2}}{\mathcal{C}^3 \pi^2} \zeta(3)^2 \simeq 
7 \times 10^{-4} \, ,
\eeq
with $\zeta (3) \simeq 1.2$.
If the right-handed neutrinos come to dominate the energy density,
they decay at a temperature
\beq
T_D^m =  \left( \frac{\pi}{4 \zeta(3)} \right)^{\frac{1}{3}} \xi^{\frac{2}{3}}
M_1 \simeq 0.87\cdot \xi^{\frac{2}{3}}  M_1 \, ,
\eeq
and assuming their decay products thermalize instantaneously, the Universe reheats to
\begin{equation}
 T^m_R = \left(\frac{\xi}{\cal C}\right)^{\frac{1}{2}} M_1 
<  \left(\frac{\xi_c}{\cal C}\right)^{\frac{1}{2}} M_1  ~.
\end{equation}
If instead, the right-handed neutrinos do not dominate the energy density and decay in a
radiation background,  the decay temperature is given by
\beq
T_D^r = \left(\frac{\xi}{\cal C}\right)^{\frac{1}{2}} M_1 \lesssim M_1 ~,
\eeq
where we have used the condition \eqref{decay}.
In this case, there is no appreciable reheating due to decay.
For later use, we define $T_{BL}$ as the temperature at which the $B-L$
asymmetry was produced (corresponding to either the decay temperature
for a radiation dominated decay, or the decay induced reheat temperature
in a matter dominated decay) or the maximum temperature when sphalerons
are in equilibrium whichever is lower. This can be regarded as the
temperature when leptogenesis occurs.

\subsection{Thermal conditions for transfer and the dark matter mass}
\label{seq:thmlcond}

The $B-L$ asymmetry generated by leptogenesis is transferred to the dark
sector via effective operators of the form
\begin{equation}
 {\cal L}_{\rm eff} = \frac{c_d}{\Lambda^{d-4}}{\cal O}_{\rm DM} {\cal O}_{\rm SM} +{\rm h.c.} ~,
\label{eq:effoptrans}
\end{equation}
if they are in thermal equilibrium, 
where ${\cal O}_{\rm DM}$ is an operator which contains only the dark
matter fields and has a non-zero dark-matter number while ${\cal
O}_{\rm SM}$ consists of SM fields only; $d\geq 4$ is the mass dimension
of the operator; $\Lambda$ denotes the scale at which the effective
operator is generated (in particular, $\Lambda = M_{\rm int}$ in the
models discussed below); $c_d$ is a dimension-less constant, which may
involve additional suppression factors such as small Yukawa couplings. 
The necessary condition for the interaction
induced by the operator to be in thermal equilibrium is then given by
\begin{equation}
 \begin{cases}
  T < T_{\rm eq} & {\rm for}~~ d = 4 \\
  T > T_{\rm eq} & {\rm for}~~ d \geq 5
 \end{cases}
~,
\end{equation}
where the decoupling temperature $T_{\rm eq}$ is determined by the
condition
\begin{equation}
 \Gamma_{{\cal L}_{\rm eff}}\bigr|_{T_{\rm eq}}
\simeq \frac{1}{8\pi^3} \frac{c_d^2}{\Lambda^{2(d-4)}}
T^{2(d-4)+1}_{\rm eq}  = \frac{{\cal C}T_{\rm eq}^2}{M_P} \simeq H\bigr|_{T_{\rm
eq}} ~,
\end{equation}
 which gives
\begin{equation}
 T_{\rm eq} \equiv \Lambda \left[\frac{8\pi^3{\cal C} \Lambda}{c_d^2
M_P}\right]^{\frac{1}{2(d-4)-1}}  ~.
\label{eq:teq}
\end{equation}
Hence, for non-renormalizable operators, if $T_{\rm eq} < T_{BL}$,
there is a period during which they are in equilibrium. If the operator is
renormalizable, then even though it is out-of-equilibrium at $T =
T_{BL}$, it will come into thermal equilibrium when
the temperature becomes lower than $T_{\rm eq}$.

The presence of such
interactions in thermal equilibrium gives rise to a condition between
the chemical potentials of SM fields and that of the dark matter field,
which relates the $B-L$ asymmetry to the asymmetry in the dark-matter
number. Let us derive this relation following the argument given in
Ref.~\cite{Ibe:2011hq}. We focus on the dominant operator in
Eq.~\eqref{eq:effoptrans}, and assume that ${\cal O}_{\rm DM}$ contains
$N_{\rm DM}$ dark matter fields (or, strictly speaking, the number of
dark matter fields minus the number of anti-dark-matter fields) and
${\cal O}_{\rm SM}$ consists of $N_Q$, $N_{u_R}$, $N_{d_R}$, $N_L$,
$N_{e_R}$, $N_H$ numbers of the left-handed quarks, right-handed up
quarks, right-handed down quarks, left-handed leptons, right-handed
charged leptons, and Higgs fields, respectively. The dark matter field
is a $n_{\rm DM}$-dimensional representation of SU(2)$_L$ and has the
hypercharge $Y_{\rm DM}$ and $B-L$ charge $Q^{\rm DM}_{B-L}$. 
By assigning each particle species a chemical potential, and 
using gauge and Higgs interactions as conditions on these potentials 
one can write down a simple set of equations for various charge densities
\cite{ht,dr}. Above the electroweak phase transition temperature, 
the conservation of the electroweak symmetry makes the chemical
potential of the $W$ boson vanish: $\mu_W=0$. In equilibrium, the
sphaleron process then yields the additional condition,\footnote{Here,
we assume that the dark matter field is either a complex scalar or a
Dirac fermion. In this case, the dark matter does not contribute to the
condition \eqref{S}. } 
\begin{equation}
  3\mu_{u_L} + \mu_{\nu_L} = 0 ~,
\label{S}
\end{equation}
where $\mu_{u_L}$ and $\mu_{\nu_L}$ are the chemical potentials
for the left-handed up quark and left-handed neutrino, respectively. 
The chemical equilibrium condition with respect to the interaction
${\cal L}_{\rm eff}$ reads
\begin{equation}
 N_{\rm DM} \mu_{\rm DM} +(N_Q + N_{u_R} + N_{d_R} ) \mu_{u_L}
+(N_L + N_{e_R}) \mu_{\nu_L} + (N_H + N_{u_R} - N_{d_R} -N_{e_R}) \mu_0
= 0 ~,
\label{eq:chemforeffop}
\end{equation}
where $\mu_{\rm DM}$ and $\mu_0$ are the chemical potentials for the
dark matter and the Higgs field. In this paper, we focus on the case
where the low-energy effective theory contains one SU(2)$_L$ doublet
Higgs boson; however, for one's convenience, in this section we keep the
number of the Higgs doublets to be arbitrary and denote it by $n_H$, with
the assumption that all of the Higgs fields have the same chemical
potential $\mu_0$.  In addition, since ${\cal L}_{\rm eff}$ should be
neutral under ${\rm U}(1)_{Y}$, we have
\begin{equation}
Y_{\rm DM} N_{\rm DM} + \frac{1}{6} N_Q + \frac{2}{3} N_{u_R} 
-\frac{1}{3} N_{d_R} -\frac{1}{2}N_L -N_{e_R} +\frac{1}{2}N_H = 0 ~.
\label{eq:leffy}
\end{equation}
On the other hand, it is not necessary for the interaction ${\cal
L}_{\rm eff}$ to conserve $B-L$ as we will see below. Let us denote the
entire $B-L$ charge of ${\cal L}_{\rm eff}$ by $\Delta_{B-L}$.
\begin{equation}
Q^{\rm DM}_{B-L} N_{\rm DM} + \frac{1}{3} N_Q + \frac{1}{3} N_{u_R} 
+\frac{1}{3} N_{d_R} -N_L -N_{e_R} = \Delta_{B-L} ~.
\label{eq:leffbl}
\end{equation}
By using Eqs.~\eqref{S}, \eqref{eq:chemforeffop}, \eqref{eq:leffy}, and
\eqref{eq:leffbl}, we then obtain\footnote{
Note that Eqs.~\eqref{eq:leffy} and \eqref{eq:leffbl} read
\begin{align}
& N_Q +N_{u_R} +N_{d_R} -3 N_L -3 N_R = 3\left[ \Delta_{B-L}
- Q^{\rm DM}_{B-L} N_{\rm DM}
\right]
 ~, \\
& N_{u_R} - N_{d_R} -N_{e_R} +N_H = Q_{B-L}^{\rm DM} N_{\rm DM}
 -\Delta_{B-L} -2 Y_{\rm  DM} N_{\rm DM} ~,
\end{align}
and  Eqs.~\eqref{S} and \eqref{eq:chemforeffop} give
\begin{align}
N_{\rm DM} \mu_{\rm DM} &= -(N_Q+N_{u_R} + N_{d_R} -3 N_L - 3 N_{e_R})
\mu_{u_L}
-(N_H + N_{u_R} - N_{d_R} -N_{e_R}) \mu_0 ~.
\end{align}
}
\begin{equation}
 \mu_{\rm DM} = 3 X_{\rm DM} \mu_{u_L} + \left(2Y_{\rm DM} -X_{\rm
					  DM}\right) \mu_{0} ~, 
\label{eq:mudmmuumu0}
\end{equation}
with 
\begin{equation}
 X_{\rm DM} \equiv Q^{\rm DM}_{B-L} - \frac{\Delta_{B-L}}{N_{\rm
  DM}} ~.
\label{eq:xdm}
\end{equation}

The electric charge density $Q$ in units of $T^2/6$ is given by
\begin{align}
 Q&= 6 \mu_{u_L} - 6 \mu_{\nu_L} + (12+2n_H) \mu_0 +2 \mu_{\rm DM}
k(z) \sum_{j=-J_{\rm DM}}^{J_{\rm DM}}
 (j+Y_{\rm DM}) \nonumber \\
&= 24\mu_{u_L} + (12+2n_H) \mu_0 +2 \mu_{\rm DM} n_{\rm DM}
 Y_{\rm DM} k(z)~,
\label{eq:qcharge}
\end{align}
where $J_{\rm DM} \equiv (n_{\rm DM}-1)/2$, $z\equiv m_{\rm DM}/T$ with
$m_{\rm DM}$ the dark matter mass, and 
\begin{equation}
 k(z) = 
\begin{cases}
 \frac{3}{4\pi^2} \int_0^{\infty} \frac{x^2 dx}{\sinh^2\left(
\frac{\sqrt{x^2 +z^2}}{2}\right)} & \text{for complex scalar} \\[3pt]
 \frac{3}{2\pi^2} \int_0^{\infty} \frac{x^2 dx}{\cosh^2\left(
\frac{\sqrt{x^2 +z^2}}{2}\right)} & \text{for Dirac fermion}
\end{cases}
~.
\end{equation}
Note that $k(z) \to 1$ for $z \to 0$, while $k(z) \propto e^{-z}$ for
$z\gg 1$. On the other hand, the dark matter multiplet does not give a
contribution to the SU(2)$_L$ charge $T_3$ due to ${\rm Tr} (T_3) = 0$. 
By using Eqs.~\eqref{eq:mudmmuumu0} and \eqref{eq:qcharge}
with the condition $Q=0$, we can express $\mu_{\rm DM}$ in terms of
$\mu_{u_L}$:
\begin{equation}
 \mu_{\rm DM} = \frac{3\left[ (10+n_H) X_{\rm DM} -8Y_{\rm DM}\right]}
{6+n_H +(2Y_{\rm DM}-X_{\rm DM})n_{\rm DM} Y_{\rm DM} k (z)} 
\mu_{u_L} ~.
\label{eq:mudm}
\end{equation}

We can also express the $B-L$ charge density in terms of
$\mu_{u_L}$. For later convenience, let us denote the contributions of
the SM and dark matter particles to the $B-L$ charge density by
$(B-L)_{\rm SM}$ and $(B-L)_{\rm DM}$, respectively, and obtain a relation
between $(B-L)_{\rm SM}$ and the asymmetry in the dark matter sector. To
that end, first we express $(B-L)_{\rm SM}$ in units of $T^2/6$ in terms
of $\mu_{u_L}$. By using Eq.~\eqref{S}, the condition $Q=0$, and
Eq.~\eqref{eq:mudm}, we 
have 
\begin{align}
(B-L)_{\rm SM} &= 3\left(4 \mu_{u_L} -3 \mu_{\nu_L} + \mu_0\right)
\nonumber \\[3pt]
&=  \frac{3\left[ 
13 n_H + 66 + 2n_{\rm DM} Y_{\rm DM} k(z) \left(13 Y_{\rm DM} -8 X_{\rm
 DM}\right) \right]}
{6+n_H +(2Y_{\rm DM}-X_{\rm DM})n_{\rm DM} Y_{\rm DM} k (z)} 
\mu_{u_L} ~.
\end{align}
Thus, the asymmetry in the dark matter sector in units of $T^2/6$,
$\Delta_{\rm DM}(z) \equiv 2 n_{\rm DM} k(z) \mu_{\rm DM}$,\footnote{We
include a factor of $n_{\rm DM}$ in the definition of $\Delta ({\rm
DM})$ since all of the charged states in the dark matter multiplet decay
into the neutral component in the end. } is related to
$(B-L)_{\rm SM}$ as 
\begin{equation}
 \Delta_{\rm DM} (z) = \frac{2 n_{\rm DM} k(z)\left[ (10+n_H) X_{\rm DM}
					     -8Y_{\rm DM}\right]}{13n_H
 + 66 +2n_{\rm DM} Y_{\rm DM} k (z) (13 Y_{\rm DM} - 8X_{\rm DM})} 
(B-L)_{\rm SM} ~.
\label{eq:deltadm}
\end{equation}
This expression shows that the interaction ${\cal
L}_{\rm eff}$ should decouple at some point; otherwise, $\Delta_{\rm
DM} (z)$ is suppressed due to the factor $k(z)$. For non-renormalizable
interactions, the decoupling temperature $T_{\rm dec}$ is equal to
$T_{\rm eq}$ if $T_{\rm eq} > m_{\rm DM}$. If $T_{\rm eq} < m_{\rm DM}$,
we need to solve the Boltzmann equation to determine the decoupling
temperature. For renormalizable interactions, once they are in thermal
equilibrium, they decouple only below the dark matter mass (or other
mass thresholds of particles participating in the interactions). 
Here, we assume that the interaction ${\cal L}_{\rm eff}$ decouples
before the electroweak phase transition. We may also consider the case
where the interaction remains in equilibrium until the time of the
electroweak phase transition, or of sphaleron decoupling. 
After decoupling, the dark matter asymmetry freezes with a value of
$\Delta_{\rm DM} \equiv \Delta_{\rm DM} (z_{\rm dec})$ where $z_{\rm eq}
\equiv m_{\rm DM}/T_{\rm dec}$.

We here note that even though the operator ${\cal L}_{\rm eff}$ violates
the $B-L$ symmetry, we can obtain a non-zero asymmetry in the dark
matter number density. This is because there is a new conserved quantity
instead of $B-L$, which makes asymmetries non-vanishing. To see this
explicitly, let us define the dark matter number $D$ for which we assign
$+1$ ($-1$) for a dark matter particle (anti-particle). The interaction
\eqref{eq:effoptrans} violates the conservation of the dark matter
number by $N_{\rm DM}$ and $B-L$ by $\Delta_{B-L}$. However, 
\begin{equation}
 X \equiv (B-L) -\frac{\Delta_{B-L}}{N_{\rm DM}} D 
\end{equation}
is conserved by the interaction. This is the new conserved quantity
which replaces $B-L$.\footnote{
Possibilities of generating non-zero baryon asymmetry in the presence of
$(B-L)$-violating interactions in equilibrium are discussed in
Ref.~\cite{krs2,bn,dr,cdeo3,dko}, where the theory possesses a conserved
quantum number which replaces $B-L$, such as lepton flavor.}

$(B-L)_{\rm SM}$ in Eq.~\eqref{eq:deltadm} is related to the baryon and lepton
asymmetries in the SM sector, $B_{\rm SM}$ and $L_{\rm SM}$,
respectively, through the ordinary procedure \cite{ht,dr}. By using
Eq.~\eqref{S} and $Q = 0$ with the dark matter contribution removed from
Eq.~\eqref{eq:qcharge}, we can express $B_{\rm SM}$ and $L_{\rm SM}$ in
terms of $\mu_{u_L}$ (in units of $T^2/6$) as  
\begin{align}
 B_{\rm SM} &= 12 \mu_{u_L}
~,\nonumber \\
 L_{\rm SM} &= 9\mu_{\nu_L} -3 \mu_0 =
-\frac{3(42 +9n_H)}{6+n_H} \mu_{u_L}
~,
\label{mus}
\end{align}
if sphaleron processes decouple before the electroweak transition. In
this case,  the relation between $(B-L)_{\rm SM}$ and $B_{\rm SM}$ is
given by
\begin{equation}
 B_{\rm SM} = \frac{4(6+n_H)}{66+13n_H} (B-L)_{\rm SM} ~.
\label{eq:bsmab}
\end{equation}
Thus, in the absence of a $B-L$ asymmetry, there is no baryon asymmetry,
and thus no dark matter asymmetry. 
If the sphaleron processes decouple after the electroweak transition \cite{latesphal}, on
the other hand, $\mu_0 = 0$ as the Higgs boson now develops a VEV, while
now $\mu_W$ is non-vanishing. In this case, the electric charge is given
by
\begin{equation}
 Q =  6 \mu_{u_L} -6\mu_{\nu_L} - 2(8+n_H) \mu_W ~.
\end{equation}
while the sphaleron condition reads
\begin{equation}
 3 \mu_{u_L} + 2\mu_W + \mu_{\nu_L} = 0 ~.
\end{equation}
Again, by imposing the electric neutrality $Q=0$, we can express $B_{\rm
SM}$ and $L_{\rm SM}$ in terms of $\mu_{u_L}$ as 
\begin{align}
 B_{\rm SM}& =  12 \mu_{u_L} + 6 \mu_W
= \frac{12(8+n_H)}{2+n_H} \mu_{u_L} ~,
\nonumber \\
L_{\rm SM}& =  9 \mu_{\nu_L} + 6 \mu_W
=-\frac{9(22 + 3n_H)}{2+n_H} \mu_{u_L}
~,
\end{align}
so that
\begin{equation}
 B_{\rm SM} = \frac{4(8+n_H)}{98+13n_H} (B-L)_{\rm SM} ~.
\label{eq:bsmbl}
\end{equation}

Provided that the symmetric part of the dark
matter sector is removed via annihilation, the present dark matter
abundance is simply given by $\Delta_{\rm DM}$. Since it is related to
$(B-L)_{\rm SM}$, we can relate it to the baryon number density today
via Eqs.~\eqref{eq:bsmab} and \eqref{eq:bsmbl}. To explain the
observed dark matter energy density, therefore, the dark matter mass
should be 
\begin{equation}
 m_{\rm DM} =m_N \left(\frac{\Omega_{c}h^2}{\Omega_Bh^2}\right)
 \left|
 \frac{13n_H
 + 66 +2n_{\rm DM} Y_{\rm DM} k (z_{\rm dec}) (13 Y_{\rm DM} - 8X_{\rm DM})}
{2n_{\rm DM} k(z_{\rm dec})\left[ (10+n_H) X_{\rm DM}
					     -8Y_{\rm DM}\right]}
\right|
\biggl[\frac{B_{\rm SM}}{(B-L)_{\rm SM}}\biggr]
 ~,
\label{eq:mdmformula}
\end{equation}
where $m_N$ is the nucleon mass.\footnote{Let us
compare our generic formulae with some results obtained in the previous
studies. Thermal conditions for the $Y_{\rm DM} = 0$ cases are presented
in Ref.~\cite{Ibe:2011hq}. Our results are consistent with Eqs.~(9) and
(10) in the published version of Ref.~\cite{Ibe:2011hq}. By setting
$n_{\rm DM} = 2$, $n_H =1$, $X_{\rm DM} = 0$, and $Y_{\rm DM} =1/2$,
Eq.~\eqref{eq:deltadm} reproduces Eq.~(8) in Ref.~\cite{Servant:2013uwa}
and Eq.~(12) in Ref.~\cite{Dhen:2015wra} with appropriate changes of
notation. }

The persistence of any $(B-L)$-violating interactions in conjunction
with electroweak sphaleron effects could wipe out \cite{fy2}
both the baryon and lepton asymmetry. Such a interaction is described by
a non-renormalizable operator which consists of only the SM fields (note
that there is no renormalizable $(B-L)$-violating operator):
\begin{equation}
 {\cal L}_{\Delta(B-L)} = \frac{c_{\Delta(B-L)}}{\Lambda^{n}} {\cal
  O}_{\Delta (B-L)} ~,
\end{equation}
where $n+4$ is the mass dimensions of the operator ${\cal O}_{\Delta
(B-L)}$, and $c_{\Delta(B-L)}$ is a constant which may include
an additional suppression factor. 
By requiring that this operator is out-of-equilibrium when
leptogenesis occurs, we obtain the following condition:
\begin{equation}
 \frac{c_{\Delta (B-L)}}{\Lambda^{n}} \lesssim 
\biggl[ \frac{8\pi^3 {\cal C} T_{BL}^{(1-2n)}}{
M_P}\biggr]^{\frac{1}{2}} ~.
\end{equation}

In the case of leptogenesis, for example, the wash-out could occur
through the $\Delta L = 2$ effective operators
of the form $y^2 LLHH / M_R$, where $y$ and $M_R$ collectively denote the
neutrino Yukawa couplings and the right-handed masses, respectively.   
This corresponds to the case where $n=1$, $\Lambda = M_R$, and
$c_{\Delta (B-L)} = y^2$.
In equilibrium, this interaction adds the condition
$\mu_{\nu_L} + \mu_0 = 0$ and hence drives all chemical potentials to 0. 
The out-of-equilibrium condition for this operator is
\beq
\Gamma_{\Delta L} = {\zeta(3) \over 8 \pi^3} {y^4 T^3 \over M_R^2} <
{\mathcal{C} T^2 \over M_P} \simeq H ~,
\eeq
yielding \cite{fy2,ht,cdeo,cdo}
\beq
{M_R \over y^2} \gtrsim 0.017 \sqrt{T_{BL} M_P} ~,
\label{fybound}
\eeq
where $H$ is the Hubble parameter.
Similarly, it is possible to put constraints on
other $B$ and/or $L$ violating operators \cite{cdeo} which
include $R$-parity violating operators in supersymmetric models.
In an inflationary context, if $M_R < m_\eta$ and the inflaton decays to right-handed neutrinos, then
it is sufficient to satisfy the constraint 
\beq
{M_R \over y^2} \gtrsim 0.017 \sqrt{{T}_R M_P} ~.
\label{fybound2}
\eeq

\subsection{Hypercharged asymmetric dark matter}

As shown in Table~\ref{cand}, some of the dark matter candidates
have a non-zero hypercharge. It is widely known that such hypercharged
dark matter is severely restricted by direct detection experiments.
Hypercharged dark matter can have a vector coupling with $Z$ boson,
which induces a spin-independent scattering with a nucleon via $Z$-boson
exchange. It turns out that its scattering cross section is larger than
the current experimental limits by orders of magnitude. 

This constraint can be evaded if there is some interaction which gives
rise to a mass splitting between the dark matter particle and its
antiparticle after electroweak symmetry breaking. In this case, a
hypercharged Dirac fermion (complex scalar) splits
into two Majorana fermions (real scalars). Since neither Majorana
fermions nor real scalars can have a vector coupling, the above
constraint can be evaded. If the mass splitting is smaller than $\sim
100$~keV \cite{Nagata:2014wma, Nagata:2014aoa}, however, inelastic
scattering via $Z$-boson exchange occurs, which is again stringently
constrained by direct detection experiments. For detailed discussions on
hypercharged dark matter in SO(10) GUTs, see Ref.~\cite{noz}.

For a Dirac fermion $\psi$ with hypercharge $Y$, the following
higher-dimensional operator can generate a mass splitting between $\psi$
and its charge conjugate $\psi^c$: 
\begin{equation}
 {\cal L}_{\Delta m} = \frac{c_{\Delta m}}{2\Lambda^{(4Y-1)} } (H^*)^{4Y}
  \overline{\psi^c} \psi +{\rm h.c.} ~,  
\label{eq:ldelm}
\end{equation}
where $H$ denotes the Higgs field, and we suppress the SU(2)$_L$
indices. $c_{\Delta m}$ is a dimension-less constant that contains the Clebsch--Gordan coefficient (see
Ref.~\cite{Nagata:2014aoa} for a more explicit expression). Once the
Higgs field develops a VEV $\langle H \rangle = v/\sqrt{2}$, this
yields a mass splitting 
\begin{equation}
 \Delta m = \frac{c_{\Delta m} v^{4Y}}{2^{(2Y-1)} \Lambda^{(4Y-1)}} ~.
\end{equation}
By requiring $\Delta m\gtrsim 100$~keV to evade the direct detection
limits, we obtain $\Lambda \lesssim 10^{9}$~GeV, $3\times 10^{4}$~GeV,
and $4\times 10^{3}$~GeV for $Y=\frac{1}{2}$, 1, $\frac{3}{2}$,
respectively, where we set $c_{\Delta m}$ equal to the Clebsch--Gordan
coefficient for the dark matter component. Similarly, for scalar dark matter with hypercharge $Y\geq
1$, the mass splitting can be induced by non-renormalizable operators
generated at a high-energy scale $\Lambda$. For $m_{\rm DM} =1$~TeV, the
requirement of $\Delta m \gtrsim 100$~keV leads to $\Lambda \lesssim
10^5$~GeV and $4\times 10^3$~GeV for $Y=1$ and $Y=\frac{3}{2}$,
respectively. For the $Y=\frac{1}{2}$ case, on the other hand, the mass
splitting is provided by a renormalizable operator, thus there is no
limit on the high-energy scale.

The operator \eqref{eq:ldelm} has the form \eqref{eq:effoptrans}, and
thus can communicate asymmetry in the SM sector to the dark matter
sector \cite{Blum:2012nf, Servant:2013uwa}. Hence, hypercharged dark
matter can be a good candidate for asymmetric dark matter, and this
possibility has been discussed in the literature
\cite{Blum:2012nf, Boucenna:2015haa, Dhen:2015wra}. As it turns out,
however, there are two challenges in this scenario, besides the direct
detection bound mentioned above. First, if the operator \eqref{eq:ldelm}
remains in thermal equilibrium below the electroweak phase transition
temperature, then it washes out the dark matter asymmetry. The chemical
equilibrium condition for this interaction gives an additional relation
between the dark matter and Higgs chemical potentials: $4Y \mu_0 + 2
\mu_{\rm DM} = 0$. After electroweak symmetry breaking, $\mu_0 = 0$,
and thus this condition implies $\mu_{\rm DM} = 0$. To avoid this, the
interaction \eqref{eq:ldelm} should decouple before electroweak
symmetry breaking. Second, the operator \eqref{eq:ldelm} causes
particle-antiparticle oscillations after electroweak symmetry
breaking, which may wash out the asymmetry in the dark sector. To
prevent this, we need to make the oscillation rate sufficiently
small or assure the decoupling of dark matter from thermal bath
before the electroweak phase transition. In the latter case, there is no
asymmetry in the dark matter sector at present, but still the
dark matter abundance is (mainly) determined by the
asymmetry of dark matter before the electroweak symmetry breaking. 

If there were no limits from direct detection experiments on the
interaction \eqref{eq:ldelm}, then we could evade these problems by
taking $\Lambda$ to be sufficiently high or the coefficient $c_{\Delta
m}$ to be very small so that the decoupling of the interaction
\eqref{eq:ldelm} is well above the electroweak transition and the
particle-antiparticle oscillation induced due to the mass splitting
$\Delta m$ is slow enough. As we will see below, however, these problems
and the direct detection bound can be evaded simultaneously only if the
dark matter mass is large enough compared with the electroweak phase
transition temperature. On the other hand, we will see in the following
discussion that in this case the annihilation of the symmetric part of
dark matter tends to be insufficient so that the dark matter relic
abundance exceeds the observed dark matter density. This incompatibility
disfavors most of the hypercharged asymmetric dark matter candidates
\cite{Blum:2012nf, Boucenna:2015haa, Dhen:2015wra}.

To see this, we give a rough estimate for the above
conditions. First, according to Eq.~\eqref{eq:teq}, $T_{\rm eq}$ for the
operator \eqref{eq:ldelm} is given by
\begin{align}
 T_{\rm eq} &= \biggl[\frac{8\pi^3{\cal C} \Lambda^{2(4Y-1)}}{c^2_{\Delta
  m} M_P}\biggr]^{\frac{1}{2(4Y-1)-1}} 
=\biggl[\frac{8\pi^3{\cal C} v^{8Y}}{4^{(2Y-1)}M_P\Delta m^2
}\biggr]^{\frac{1}{2(4Y-1)-1}} ~.
\end{align}
For $Y=1/2$, for instance, this reads
\begin{equation}
 T_{\rm eq} \simeq 100~{\rm GeV} \times \left(\frac{100~{\rm
					 keV}}{\Delta m}\right)^2 ~.
\end{equation}
This result shows that the requirement $\Delta m \gtrsim 100$~keV to
evade the direct detection bound may cause the operator \eqref{eq:ldelm}
to remain
in equilibrium down to the electroweak phase transition. We however note
that the formula \eqref{eq:teq} is based on the assumption that all of
the relevant particles are relativistic. Thus, if the dark matter mass
is much larger than the electroweak scale, the above consequence may be
modified significantly.

The second condition follows from $\Gamma_{\rm osc} < H|_{T_{\rm EW}}$
where $\Gamma_{\rm osc} = \Delta m/2$ is the rate of particle-antiparticle
oscillations and $T_{\rm EW}$ is the temperature at the electroweak
phase transition. This leads to
\begin{equation}
 \Delta m < \frac{2{\cal C} T_{\rm EW}^2}{M_P} \simeq 3 \times 10^{-14}
  ~{\rm GeV} \times \left(\frac{T_{\rm EW}}{100~{\rm GeV}}\right)^2 ~.
\end{equation}
Obviously, this conflicts with the direct detection bound. Thus, to avoid
particle-antiparticle oscillations from erasing the dark matter
asymmetry, the dark matter should decouple from thermal bath above
$T_{\rm EW}$. Since the freeze-out temperature of dark matter is given
by $\simeq m_{\rm DM}/25$, this condition requires $m_{\rm DM} \gtrsim
25 T_{\rm EW}$.

As we have just seen, the above conditions may be evaded if $m_{\rm DM}
\gg T_{\rm EW}$. On the other hand, there is an upper bound on the dark
matter mass which follows from the requirement that the symmetric part
of dark matter be annihilated away so that the asymmetric part accounts
for the (dominant part of the) dark matter abundance. For example, for the
SU(2)$_L$ doublet $Y=1/2$ Dirac dark matter, the annihilation is effective if
$m_{\rm DM} < 1$~TeV \cite{Cirelli:2007xd}. On the other hand, the
second condition discussed above requires $m_{\rm DM} \gtrsim 25 T_{\rm
EW} > 1$~TeV, and thus the doublet Dirac fermion is unable to be
asymmetric dark matter \cite{Blum:2012nf}. For the SU(2)$_L$
doublet scalar dark matter, the upper bound on the dark matter mass is
relaxed if the dark matter-Higgs quartic coupling is large. Even in this case,
however, the dark matter asymmetry is found to be much smaller than
the observed dark matter density once the perturbativity condition is
imposed on the quartic coupling \cite{Dhen:2015wra}. Other cases for
hypercharged dark matter candidates are discussed in
Ref.~\cite{Boucenna:2015haa}, and found that the $Y > 1$ cases are
excluded. As a consequence, only the $Y=1$ cases can be promising candidates
for hypercharged asymmetric dark matter.

\subsection{Candidate models for SO(10) asymmetric dark matter}

Let us summarize the discussion in this section, and list up promising
candidates for asymmetric dark matter in SO(10) GUTs. First, we consider
the ``minimal models'', namely, we require that besides the SM particles
only the dark matter multiplet has a mass much lighter than the
intermediate scale. In this case, the low-energy effective theory
only contains the SM particles and the dark matter, and the relevant
non-renormalizable operators are generated at the intermediate or GUT
scale.

As discussed in the previous subsection, the $Y=1/2$ and $3/2$
candidates in Table~\ref{cand} have already been excluded. 
In addition, the analysis in Ref.~\cite{noz} shows that
\model{S}{3}{0}, \model{S}{3}{1}, \model{F}{3}{1},
\model{\widehat{F}}{1}{0}, \model{\widehat{F}}{3}{1} are not good
candidates for SO(10) dark matter models. This is because none of these models 
are consistent with gauge coupling unification with reasonable values of $M_{\rm int}$ and/or
$M_{\rm GUT}$ with minimal field content. As a result, only
\model{S}{1}{0} can be a promising candidate for SO(10) asymmetric dark
matter. We will discuss this candidate in the subsequent
section. Then, we discuss some next-to-minimal extensions in
Sec.~\ref{sec:NMmodels}.

\section{Scalar Singlet Asymmetric Dark Matter}
\label{sec:scalarsinglet}

As we discussed in the previous section, singlet scalar dark matter is
the only candidate for asymmetric dark matter in SO(10) if we require
the minimality. We discuss this possibility in this section. First, in
Sec.~\ref{sec:scasingdmmodels}, we summarize possible scalar singlet
dark matter models in SO(10) following the discussion in
Ref.~\cite{noz}. In Sec.~\ref{sec:anni}, we derive the condition that
the symmetric part of the scalar singlet dark matter is sufficiently
annihilated away, and discuss the current experimental constraints. 
In Sec.~\ref{sec:oscillations}, we argue for the
necessity of suppressing the particle-antiparticle oscillations, and
show that we can actually evade the oscillations by taking some relevant
Lagrangian terms to be very small. Then, in Sec.~\ref{sec:transfer}, we discuss
the case where the dark matter asymmetry is thermally transferred from
the $B-L$ asymmetry in the SM sector. We demonstrate that this
possibility does not work for the SO(10) dark matter candidates given in
Sec.~\ref{sec:scasingdmmodels}. Finally, we consider the cogenesis
scenario in Sec.~\ref{sec:cogenesis}.

\subsection{Scalar singlet dark matter candidates}
\label{sec:scasingdmmodels}

As shown in Table~\ref{cand}, singlet scalar dark matter belongs to
either a ${\bf 16}$ or ${\bf 144}$ of SO(10). Below the GUT scale, there
are several possibilities for the dark matter multiplet according to
different intermediate gauge groups. As shown in Ref.~\cite{noz}, among
these possibilities, only three accommodate a sufficiently high GUT
scale, which is required to evade the proton decay bound. These models
are called \DM{SA}{422}, \DM{SA}{3221}, and \DM{SA}{3221D} in
Ref.~\cite{noz}, where the intermediate gauge groups are $\text{SU}(4)_C
\otimes \text{SU}(2)_L \otimes \text{SU}(2)_R$, $\text{SU}(3)_C
\otimes \text{SU}(2)_L \otimes \text{SU}(2)_R \otimes
\text{U}(1)_{B-L}$, and $\text{SU}(3)_C \otimes \text{SU}(2)_L \otimes
\text{SU}(2)_R \otimes \text{U}(1)_{B-L} \otimes D$, respectively, with
$D$ denoting the so-called $D$-parity \cite{Kuzmin:1980yp}. In the case
of \DM{SA}{422}, we fine-tune the mass of the ({\bf 4}, {\bf 1}, {\bf 2})
representation of $\text{SU}(4)_C \otimes \text{SU}(2)_L \otimes
\text{SU}(2)_R$ in the dark matter multiplet to be much lighter than the
GUT scale, while the rest of the components remain around the GUT
scale. After the intermediate gauge symmetry is broken, only the singlet
complex scalar component of the ({\bf 4}, {\bf 1}, {\bf 2}) has a mass
around the TeV scale via another fine-tuning, while the other components
remain with masses of ${\cal O}(M_{\rm int})$. By solving RGEs, we obtain
$M_{\rm GUT} =2.1\times 10^{16}$~GeV and $M_{\rm int} = 1.2 \times
10^{11}$~GeV, and the proton lifetime is computed to be $\tau (p\to
e^{+} \pi^0) = 6.4\times 10^{36}$~yrs, where the SO(10) gauge boson mass
is set equal to $M_{\rm GUT}$. Similarly, for \DM{SA}{3221}, the $({\bf
1}, {\bf 1}, {\bf 2}, 1)$ component lies below the GUT scale. The GUT
and intermediate scales are found to be $4.6\times 10^{16}$~GeV and
$3.4\times 10^{8}$~GeV, respectively, and the proton lifetime is
$1.4\times 10^{38}$~yrs. For \DM{SA}{3221D}, in addition to the $({\bf
1}, {\bf 1}, {\bf 2}, 1)$, the $({\bf 1}, {\bf 2}, {\bf 1}, 1)$
component also lies around the intermediate scale due to
$D$-parity. In this case, we have $M_{\rm GUT} = 3.8\times 10^{15}$~GeV,
$M_{\rm int} = 1.2\times 10^{10}$~GeV, and $\tau (p \to e^+ \pi^0) =6.7
\times 10^{33}$~yrs. We note that the SO(10) gauge boson masses can be
different from $M_{\rm GUT}$ by an ${\cal O}(1)$ factor, and thus we
expect an order of magnitude uncertainty in the computation of proton
lifetimes. Taking this uncertainty into account, all of these models are
consistent with the present proton decay bound $\tau (p \to e^+ \pi^0) >
1.6 \times 10^{34}$~yrs \cite{Miura:2016krn}.

To be specific, we focus on the \DM{SA}{3221} case in the following
analysis, but similar discussions can also be applied to the other cases.

\subsection{Annihilations and experimental limits}
\label{sec:anni}

\begin{figure}[t]
\begin{center}
\includegraphics[width = 0.7 \textwidth]{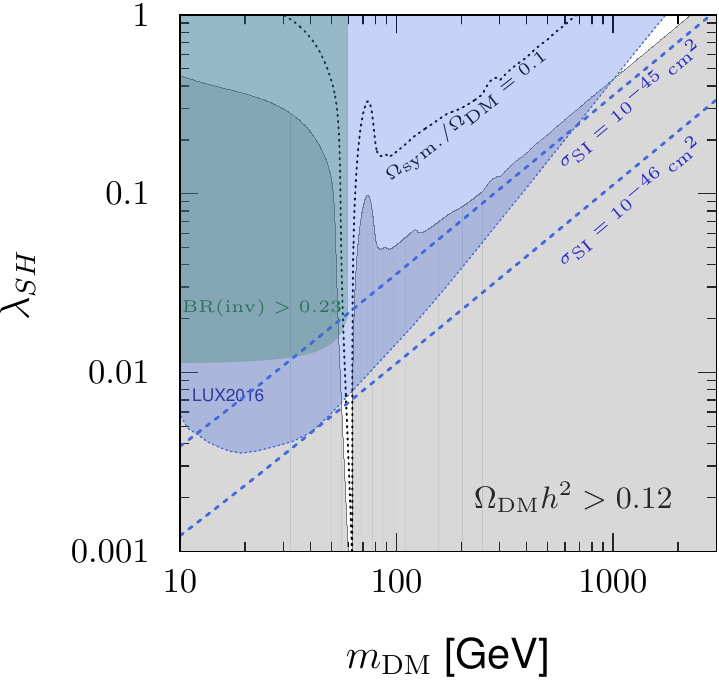}
\caption{{\it  Constraints on the complex scalar singlet dark matter. The
 gray shaded area is excluded since the predicted dark matter abundance
 exceeds the observed value $\Omega_{\rm DM} h^2 = 0.12$
 \cite{Planck}. The green shaded region is excluded since the invisible
 decay branch of the Higgs boson is too large. The blue shaded region is
 excluded by the LUX 2016 result \cite{Akerib:2016vxi}. The black dotted
 line shows the parameter points where the symmetric part is 10\% of the
 observed dark matter density. The upper (lower) blue dashed line
 corresponds to $\sigma_{\rm SI}^{(p)} = 10^{-45}~{\rm cm}^2$
 ($10^{-46}~{\rm cm}^2$). }}
\label{fig:limit}
\end{center}
\end{figure}

In order for the asymmetric dark matter scenario to work, the symmetric
part of dark matter should efficiently be eliminated. This requirement
gives a lower bound on the annihilation rate of dark matter. The
annihilation of the singlet dark matter in our model proceeds through the
dark matter-Higgs quartic coupling:
\begin{equation}
 {\cal L}_{\rm int} = - \lambda_{SH} |S|^2 |H|^2 ~,
\label{eq:lamsh}
\end{equation}
where $S$ denotes the scalar singlet dark matter. For a dark matter mass
smaller than the weak gauge boson masses, the dominant annihilation mode
is $S S^* \to b\bar{b}$. For $m_{\rm DM} > \text{few}\times 100$~GeV, on
the other hand, the dark matter particles annihilate into a pair of the SM Higgs
bosons or weak gauge bosons. We compute the relic abundance of the
symmetric part, $\Omega_{\rm sym.} h^2$, using {\tt micrOMEGAS}
\cite{Belanger:2014vza}. We need to require that $\Omega_{\rm sym.} h^2$
is much smaller than the observed dark matter density, $\Omega_{\rm
sym.} h^2 \ll 0.12$ \cite{Planck}, in order for the asymmetric part to
be the dominant component of the dark matter abundance. In
Fig.~\ref{fig:limit}, we show in the $m_{\rm DM}$--$\lambda_{SH}$
parameter space the region where $\Omega_{\rm sym.} h^2 > 0.12$ in the
gray shaded area, which is phenomenologically excluded. In addition, the
black dotted line shows the parameter points where the symmetric part is
10\% of the total dark matter density. The region above (below) the line
predicts a smaller (larger) abundance for the symmetric part. As we can
see, $\lambda_{SH} \gtrsim 0.1$ is required to sufficiently remove the
symmetric part, with the exception of the resonance region $m_{\rm DM} \simeq m_h/2$ 
with $m_h\simeq 125$~GeV the mass of the Higgs boson
\cite{Aad:2015zhl}.

If the mass of the dark matter singlet is smaller than $m_h/2$, 
the SM Higgs boson can decay into a pair of the dark matter particles
through the interaction \eqref{eq:lamsh}. This decay mode is invisible
at the LHC, and reduces the branching fractions of the other decay
channels, which is severely restricted by the Higgs measurements at the
LHC \cite{Khachatryan:2016vau}. Currently, LHC experiments give an
upper limit on the branching ratio of the Higgs invisible decay mode,
${\rm BR}({\rm inv.}) < 0.23$ \cite{Aad:2015pla, CMS:2016rfr}. In our
model, the branching ratio of the invisible decay is evaluated as
\begin{equation}
 {\rm BR}({\rm inv.}) = \frac{\Gamma ( h\to SS^*)}{\Gamma (h \to SS^*) +
  \Gamma_{\rm Higgs}} ~,
\end{equation}
where $\Gamma_{\rm Higgs} = 4.07 \times 10^{-3}$~GeV is the total decay
width of a 125~GeV Higgs boson, and 
\begin{equation}
 \Gamma (h \to SS^*) = \frac{\lambda_{SH}^2 v^2}{16\pi m_h}
\sqrt{1-\frac{4 m_{\rm DM}^2}{m_h^2}} ~.
\end{equation}
Thus, the upper limit on ${\rm BR}({\rm inv.})$ leads to an upper limit
on $\lambda_{SH}$ for $m_{\rm DM} < m_h/2$. We show the region excluded
by the Higgs invisible decay bound in the green shaded region in
Fig.~\ref{fig:limit}. This indicates that $m_{\rm DM} < m_h/2$ 
is excluded by a combination of this bound and the relic density of the symmetric part.

The interaction \eqref{eq:lamsh} also induces dark matter-nucleon
scatterings via Higgs boson exchange, and thus direct detection
experiments impose limits on the coupling $\lambda_{SH}$.\footnote{In
the low dark matter mass region $m_{\rm DM} \lesssim 10$~GeV, the dark
matter-nucleon scattering cross section is also restricted by the
existence of neutron stars \cite{McDermott:2011jp}. } The
spin-independent (SI) scattering cross section of dark matter with a
nucleon $N$ ($N = p$ or $n$) is computed as follows \cite{Shifman:1978zn}:
\begin{equation}
 \sigma_{\rm SI}^{(N)} = \frac{f_N^2 m_N^2}{4\pi (m_N + m_{\rm DM})^2} ~,
\end{equation}
where $m_N$ is the nucleon mass and $f_N$ is given by
\begin{equation}
 \frac{f_N}{m_N} = \frac{\lambda_{SH}}{m_h^2}
\left[\sum_{q=u,d,s} f_{T_q}^{(N)} +\frac{2}{9} f_{TG}^{(N)}\right] ~,
\end{equation}
with the mass fractions $f_{T_q}^{(N)} \equiv \langle N | m_q \bar{q}q
|N\rangle/m_N$, $m_q$ being the mass of a light quark $q$, and $f_{TG}^{(N)} \equiv
1-\sum_{q=u,d,s} f_{T_q}^{(N)}$. For the mass fractions we use the
values computed by using lattice simulations in
Ref.~\cite{Abdel-Rehim:2016won}: $f_{T_u}^{(p)} = 0.0149$,
$f_{T_d}^{(p)} =0.0234$, and $f_{T_s}^{(p)} = 0.0440$. If we instead use
the values obtained  from the pion-nucleon $\sigma$-term $\Sigma_{\pi
N}=50$~MeV \cite{Ellis:2008hf} and $\sigma_0 = 36$~MeV
\cite{Borasoy:1996bx} following the discussion given in
Ref.~\cite{Ellis:2008hf}, we obtain a larger SI scattering cross section
by a factor of $\simeq 2$. We show the contour of $\sigma_{\rm SI}^{(p)}
= 10^{-45}~{\rm cm}^2$ ($10^{-46}~{\rm cm}^2$) by the upper (lower) blue
dashed line in Fig.~\ref{fig:limit}. We also show the current constraint
from the LUX experiment \cite{Akerib:2016vxi} by the blue shaded area. We
find that there are two allowed regions: $m_{\rm DM} \simeq m_h/2$ and
$\simeq 1$~TeV. In the latter case, the symmetric part is the dominant
component of the dark matter abundance. For a larger dark matter mass, a
larger $\lambda_{SH}$ is required to eliminate the symmetric part of the
dark matter abundance. However, $\lambda_{SH}$ cannot be too large; the
requirement of perturbativity up to the GUT scale imposes an upper bound
on the value of $\lambda_{SH}$ at the electroweak scale
\cite{mnoz}. This upper bound depends on the particle content and other
couplings in the model, but typically $\lambda_{SH} \lesssim
1$ \cite{mnoz}. According to Fig.~\ref{fig:limit}, this indicates
$m_{\rm DM} \lesssim 2$~TeV. This dark matter mass region will soon be
tested in dark matter direct detection experiments such as XENON1T
\cite{Aprile:2015uzo}. In the former case, on the other hand, although
it is possible for the asymmetric part to dominate the symmetric part,
it is hard to probe the whole parameter space in future experiments due
to the fact that $\lambda_{SH}$ must be small \cite{Abe:2014gua}.

\subsection{Particle-antiparticle oscillations}
\label{sec:oscillations}

Since $S$ is a singlet under the SM gauge group, in addition to the
particle-number-conserving mass term $|S|^2$, it can also have the
particle-number-violating mass terms $S^2$ and $S^{*2}$. These mass
terms induce particle-antiparticle oscillations $S\leftrightarrow
S^*$, which are problematic as they erase the asymmetry in the dark
matter sector. To avoid this problem, the oscillation rate has to be
small, namely, 
\begin{equation}
 \Gamma_{\rm osc} = \frac{\Delta m}{2} \lesssim \frac{{\cal C}}{M_P} 
\biggl(\frac{m_{\rm DM}}{25}\biggr)^2 ~,
\label{eq:consosci}
\end{equation}
where $\Delta m $ denotes the mass splitting between the dark matter
particle and its antiparticle induced by the particle-number-violating
mass terms, and the right-hand side is the Hubble parameter when the
dark matter decouples from the thermal bath. In the presence of the mass
terms $\mu^2 (S^2+S^{*2})/2$, the mass splitting is given by $\Delta m =
\mu^2/m_{\rm DM}$, and thus Eq.~\eqref{eq:consosci} leads to
\begin{equation}
 \mu \lesssim \frac{m_{\rm DM}}{25} \biggl(\frac{2 {\cal C} m_{\rm
  DM}}{M_P} \biggr)^{\frac{1}{2}} \simeq
2\times 10^{-6}
\times\biggl(\frac{m_{\rm DM}}{1~{\rm TeV}}\biggr)^{\frac{3}{2}} 
~{\rm GeV}
~.
\label{eq:boundonmu}
\end{equation}

In SO(10), both ${\bf 16}^2$ and ${\bf 144}^2$ are forbidden by the
gauge symmetry, and thus the particle-number-violating mass terms are
absent. The intermediate gauge symmetries also forbid such mass
terms. Below the intermediate gauge scale, however, the VEV of the {\bf
126} Higgs field can generate the particle-number-violating mass terms via
the interactions
\begin{equation}
 {\cal L}_{\rm int} = - \kappa_2 R_{\rm DM} R_{\rm DM} R_2^* 
-\lambda_{12}^{\bf 126} (R_{\rm DM} R_{\rm DM})(R_1
R_2^*)_{\overline{\bf 126}} +{\rm h.c.}~,
\end{equation}
where $R_{\rm DM} = {\bf 16}$ or ${\bf 144}$ denotes the dark matter
multiplet, $R_1$ is the GUT Higgs field, $R_2 ={\bf 126}$ is the
intermediate-scale Higgs field, and the subscripts after the parentheses
denote the SO(10) representation formed by the product in them. To
satisfy the bound \eqref{eq:boundonmu}, we need to suppress the
couplings $\kappa_2$ and $\lambda_{12}^{\bf 126}$. Once they are taken
to be small, they remain small under the renormalization flow. By
making these couplings small, we can also suppress the
particle-number-violating couplings with the Higgs boson, such as
$S^2|H|^2$, which are induced by the exchange of the {\bf 126} Higgs and
lead to the particle-number-violating mass terms after electroweak
symmetry breaking.

\subsection{Thermal transfer}
\label{sec:transfer}

If the asymmetry in the dark sector is transferred from the $B-L$
asymmetry in the SM sector through effective interactions
\eqref{eq:effoptrans}, the dark matter mass is uniquely determined by
the thermal relation \eqref{eq:mdmformula}. The lowest-dimension
effective operator which has the form of \eqref{eq:effoptrans} is 
\begin{equation}
 {\cal L}^{(7)}_{\rm eff} = \frac{c_7}{\Lambda^3} S^2 H^2 \overline{L^c}
  L +{\rm h.c.} ~,
\end{equation}
which can be induced by the exchange of the intermediate-scale
particles. We thus take $\Lambda = M_{\rm int}$ with other possible
suppression factors included in the coefficient $c_7$. $T_{\rm eq}$ for
this operator is then given by Eq.~\eqref{eq:teq}:
\begin{equation}
 T_{\rm eq} = M_{\rm int} \biggl[\frac{8\pi^3 {\cal C} M_{\rm int}}
{c_7^2 M_P}\biggr]^{\frac{1}{5}} ~.
\end{equation}

If $T_{\rm eq} \gg m_{\rm DM}$, this interaction decouples from the thermal
bath much before the decoupling of the dark matter, and in particular we
can set $k(z_{\rm dec}) = 1$ in Eq.~\eqref{eq:mdmformula}. By setting
$n_{\rm DM} = 1$, $n_H = 1$, $Y_{\rm DM} = 0$, and $X_{\rm DM} = Q^{\rm
DM}_{B-L} = 1$, we then obtain
\begin{equation}
 m_{\rm DM} = m_N \frac{79}{22}
\biggl(\frac{\Omega_c h^2}{\Omega_B h^2}\biggr)
\biggl[\frac{B_{\rm SM}}{(B-L)_{\rm SM}}\biggr]
\simeq 6.0~{\rm GeV} ~,
\end{equation}
where we have used Eq.~\eqref{eq:bsmbl}. However, such a small dark
matter mass has already been excluded by the constraint on the Higgs
invisible decay width as shown in Fig.~\ref{fig:limit}. 

If $T_{\rm eq} \lesssim m_{\rm DM}$, on the other hand, the dark matter
mass given by Eq.~\eqref{eq:mdmformula} can be increased due to the
Boltzmann factor $k(z_{\rm dec})$. In terms of the intermediate scale
$M_{\rm int}$, the inequality $T_{\rm eq} \lesssim m_{\rm DM}$ reads
\begin{equation}
 M_{\rm int} \lesssim m_{\rm DM}^{\frac{5}{6}} \biggl[\frac{c_7^2 M_P}
{8\pi^3 {\cal C}}\biggr]^{\frac{1}{6}} 
\simeq
c_7^{\frac{1}{3}} \times \biggl(\frac{m_{\rm DM}}{1~{\rm
TeV}}\biggr)^{\frac{5}{6}} \times 10^5~{\rm GeV} ~.
\label{lowint}
\end{equation}
As discussed in Sec.~\ref{sec:scasingdmmodels}, however, there is no
candidate in a model with minimal field content
which predicts such a low intermediate scale.\footnote{We note that one can construct a non-minimal
model with a low intermediate scale. This can be done for example, if the intermediate gauge group is broken in two steps to the SM.  While one of the intermediate scales remains relatively large, the 
second may be as low as $\sim 1$ TeV \cite{Evans:2015cqq}.} We therefore
conclude that the thermal transfer scenario does not work for the scalar
singlet asymmetric dark matter candidate in SO(10).

\subsection{Cogenesis}
\label{sec:cogenesis}

Next, we discuss the cogenesis scenario. In this case, the asymmetry in
the dark matter sector is directly produced via the decay of
intermediate-scale particles and does not communicate to the SM sector
after the generation of the asymmetry. To that end, for instance, we 
may utilize the decay of right-handed neutrinos by introducing a new
singlet fermion $\psi_S$ around the intermediate scale. The singlet
fermion can have a coupling with the scalar singlet dark matter and
right-handed neutrinos if the singlet fermion is in a {\bf 1}, {\bf 45},
or {\bf 210} of SO(10).\footnote{If $R_{\rm DM} = {\bf 144}$, a {\bf 54}
is also possible. } For brevity, we take it to be a {\bf 1} in the
following discussion. Such a singlet field does not affect the running
of gauge couplings, and thus does not spoil gauge coupling
unification. If the mass of the new singlet fermion is smaller than the
masses of right-handed neutrinos, the right-handed neutrinos can decay
into the singlet fermion and the scalar singlet dark matter. If the new
couplings contain CP phases, the dark matter asymmetry is produced in
the decay process just like leptogenesis. The singlet fermion can decay
into the dark matter and SM particles via the exchange of right-handed
neutrinos afterward. Although this decay process modifies the primordial
asymmetry in the dark sector, it does not completely erase the asymmetry
in general since the asymmetry generated at the late time depends on a
different combination of the Yukawa couplings from that for the
primordial asymmetry. 

Alternatively if the mass of $\psi_S$ is greater than $m_{\nu_R}$,
then if $\psi_S$ is produced during reheating, its out-of-equilibrium decays
will produce an asymmetry in the dark matter scalar $S$. The out-of-equilibrium
decay of $\nu_R$ produces a lepton asymmetry as in ordinary leptogenesis.
In either case, 
the net dark matter asymmetry depends on the
branching fractions of right-handed neutrinos and the new CP phase, but
generically it is the same order as the $B-L$ asymmetry in the SM
sector, which may explain why $\Omega_c h^2$ and $\Omega_B h^2$ are the
same order of magnitude.

Though the cogenesis scenario may work in this case, as shown in
Fig.~\ref{fig:limit}, the allowed dark matter mass region is stringently
restricted and can be soon tested in future experiments (except for the
very narrow region around $m_{\rm DM} \simeq m_h/2$). This motivates us
to move to the next-to-minimal models, which will be discussed in the
subsequent section.

\section{Next-to-minimal models}
\label{sec:NMmodels}

In the models we considered in the previous sections, the dark matter
particle develops an asymmetric part in its density, either through
transfer from an asymmetry of the SM particles or by cogenesis, and
preserves it as (a part of) the dark matter relic observed today. As we have seen, these
models are severely constrained leaving only the scalar singlet dark matter model of cogenesis
with its mass limited to a narrow window around $m_h/2$, (the window
around $\sim 1$~TeV is also possible, though in this case most of the dark matter
abundance originates from the ordinary thermal relic). However, we may
find additional models if we relax the notion of the asymmetric dark
matter---namely, the constraints discussed above can be relieved if the
dark matter relic abundance is only required to have an asymmetric
origin while it can be totally symmetric today. We discuss this
possibility in this section.

More specifically, we consider dark matter models that achieve the relic
density in two steps, similar to models considered in
Refs.~\cite{Servant:2013uwa, Dhen:2015wra}. In these models, two
$\mathbb{Z}_2$-odd particles (or multiplets) $X_1$ and
$X_2$ are introduced near the TeV scale. $X_1$ is the lighter one whose
relic density eventually originates from the $B-L$ asymmetry in two steps:
i) $X_2$ obtains asymmetric density either by cogenesis or by asymmetry
transfer from SM particles, and then ii) the asymmetric density in
$X_2$ is converted to the relic density of $X_1$ through $X_2$ decay. In
order to annihilate the symmetric part of the thermal abundance
efficiently, $X_1$ needs to have sizable couplings with the SM
sector. To that end, we assume that $X_1$ has a charge under the
$\text{SU}(2)_L \otimes \text{U}(1)_Y$ gauge interactions. In the
models we present below, $X_1$ is a Majorana fermion.  
$X_2$ needs to have a long enough lifetime to decay after depletion of
$X_1$ symmetric density; otherwise the determination of $X_1$ relic
density is similar to that in the usual thermal relic scenario. 

In the rest of this section we will present two non-minimal models where
the asymmetry of $X_2$ is acquired by cogenesis and by transfer,
respectively. In Sec.~\ref{sec:ExtCogenesis}, we extend the cogenesis
model discussed in Sec.~\ref{sec:cogenesis} so that the scalar singlet
plays the role of $X_2$ and the dark matter particle $X_1$ is a
singlet-doublet Majorana particle. In Sec.~\ref{sec:stop-bino}, we
consider a model where $X_2$ is a stop-like particle whose asymmetry is
transferred from the top quark by a Yukawa interaction.

\subsection{Extension of cogenesis model}
\label{sec:ExtCogenesis}

In the model considered in
Sec.~\ref{sec:cogenesis}, the asymmetry in the dark matter sector can be
generated by the lightest right-handed neutrino decay
$N_1\rightarrow S+\psi_S$. In addition, we introduce an additional
$\mathbb{Z}_2$-odd Dirac doublet $\psi_D$ around the TeV scale. The
neutral component of the doublet $\psi_D^0$ is mixed with the fermionic
singlet $\psi_S$ through the Yukawa coupling 
\begin{equation}
\mathcal{L}_{\rm Yukawa} = - \kappa \psi_S {\psi_D} H + {\rm h.c.}
\end{equation}
after the Higgs field develops a VEV. Furthermore, we choose the masses
of $\psi_S$, $\psi_D$, and $S$---$m_{\psi_S}$, $m_{\psi_D}$, and $m_S$,
respectively---such that $m_{\psi_S}, m_{\psi_D} < m_S < M_1$. The lightest 
$\mathbb{Z}_2$-odd particle is a $\psi_D^0-\psi_S$ mixture which we
denote by $\chi$ in the following text becomes the dark matter candidate. 

There are two ways for $\chi$ to achieve the correct dark matter
relic density. The first possibility is that $S$ decays early into
$\chi$, and then $\chi$ annihilates through its $\psi_D^0$ component by a
gauge interaction. In this case, the primordial asymmetry generated in
the dark sector is washed out since $\chi$ is a Majorana particle, and
thus this case is just the ordinary thermal relic scenario. Here, we
consider the other possibility; both $S$ and $\chi$ nearly fully annihilate 
their symmetric density before the asymmetric part of $S$ decays
into $\chi$. In this case, the annihilation cross sections of both $S$
and $\chi$ are larger than those required in the thermal relic scenario
so that the symmetric part of their abundance is suppressed after the
annihilations freeze out.

By requiring a low symmetric relic density, we can constrain the 
masses of $S$ and $\chi$. The relic abundance of an SU(2)$_L$ doublet
Dirac dark matter candidate is saturated by the symmetric part if its
mass is about $1~{\rm TeV}$ \cite{Cirelli:2007xd}. Thus, 
if we require that the density of asymmetric origin makes up over $90\%$
of the total relic density, we can set a bound on the dark matter particle
mass $m_{\chi} \lesssim 1{~\rm TeV}/\sqrt{10}\sim 350 {\rm GeV}$. The
DM-nucleon scattering cross section for almost pure SU(2)$_L$
doublet dark matter is found to be very small ($\sigma_{\rm SI} \lesssim
10^{-49}~{\rm cm}^2$) \cite{Hisano:2015rsa} and thus this candidate
can evade the direct detection limits.\footnote{If $\chi$ is a well-mixed state of
singlet and doublet components, the dark matter-nucleon scattering is
induced by the Higgs boson exchange process, which is severely
constrained by the direct detection experiments \cite{Mahbubani:2005pt,
Banerjee:2016hsk}. However, there is a specific parameter region,
so-called blind spot \cite{Falk:1998xj, cancel, Cheung:2012qy}, where
the direct detection bound is evaded even though the singlet-doublet
mixing is sizable. In this region, the symmetric part of dark matter
relic agrees with the observed dark matter density even if the dark matter
mass is as large as $\sim 1.5$~TeV \cite{Banerjee:2016hsk}; therefore,
for the symmetric origin of the dark matter abundance to be less than
10\%,  $m_{\rm DM}\lesssim 1.5{~\rm TeV}/\sqrt{10}\sim 500~{\rm GeV}$ is
required in the case of the blind spot. 
}
We recall from Fig.~\ref{fig:limit} that $m_S\lesssim 700~{\rm GeV}$
if its symmetric density contributes less than $10\%$ of the dark matter
relic density and if we assume perturbativity of $\lambda_{SH}$.

The decay of $S$ proceeds via its coupling to the matter {\bf 16} and the singlet in $\chi$ so that $S$
to $\chi$ is mediated by right-handed neutrinos: 
$S\rightarrow \chi+N_i^{(*)}\rightarrow \chi+L_i+H$, where $N_i^{(*)}$
represent the virtual intermediate $N_i$ and $L_i$ are the left-handed
lepton doublets. The decay width of $S$ is estimated as 
\begin{equation}
 \Gamma_S \simeq \sum_{i}\frac{\lambda_i^2 y_i^2}{3 \times 2^8 \pi^3}
  \frac{m_S^3}{M_i^2} ~,
\label{eq:gammas}
\end{equation}
where $\lambda_i$ denote the $\chi$--$S$--$N_i$ couplings and $y_i =
y_{ii}$. If $m_S \simeq m_{\chi}$, then this decay width is further
suppressed by a phase space factor. Now suppose that the exchange of the
lightest right-handed neutrino $N_1$ dominates the scattering
amplitude. Here, we note that the relevant couplings $\lambda_1$ and
$y_1$ are restricted by the out-of-equilibrium decay condition
\eqref{decay}: $(\lambda_1^2 + y_1^2) \lesssim 8\pi {\cal C}
M_1/M_P$. This then gives
\begin{align}
 \Gamma_S \lesssim \frac{1}{3 \times 2^8 \pi^3} \frac{m_S^3}{M_1^2} 
\times \frac{4^2 \pi^2 {\cal C}^2 M_1^2}{M_P^2}
=\frac{{\cal C}^2 m_S^3}{48\pi M_P^2} ~,
\end{align}
and thus a lower limit on the lifetime of $S$, $\tau_S$, is obtained as 
\begin{equation}
 \tau_S \gtrsim 4\times 10^5 \times \left(\frac{500~{\rm
				     GeV}}{m_S}\right)^3 
~{\rm s}~.
\end{equation}
Such slow decay can re-process the light element abundances produced by
the Big-Bang nucleosynthesis (BBN), and thus is strongly constraint by
the success of BBN calculation. It is shown in Ref.~\cite{Cyburt:2009pg}
that the BBN constraint starts to come into play when $\tau_S\gtrsim
100{~\rm s}$. To make $\tau_S< 100{~\rm s}$, the decay of $S$ should be
dominated by the exchange of heavier right-handed neutrinos, since the
Yukawa couplings for these heavier right-handed neutrinos, $y_2$ and
$y_3$, are in general not limited by the condition \eqref{decay} since there is no reason 
to impose that $N_{2,3}$ decay out-of-equilibrium. Then the decay of $S$ can be dominated
by $N_{2,3}$ 
by assuming $\lambda_{2,3} \gg \lambda_1$.   We
however note that it is necessary to ensure that the asymmetry wash-out
scatterings $LL\leftrightarrow H^*H^*$ and
$SS\leftrightarrow\psi_S\psi_S$ decouple at the temperature $T_{BL}$
defined in Sec.~\ref{sec:genofasym}. Suppose the exchange of $N_i$,
$i\neq 1$ dominates the $S$ decay. Then, let us estimate the bound
on $y_i$ and $\lambda_i$ coming from the decoupling conditions of the
wash-out processes; this is given by Eq.~\eqref{fybound} as
$y_i^2, \lambda_i^2 \lesssim M_i/(0.017\sqrt{T_{BL} M_P})$.   
The decay rate of $S$ is then 
\begin{align}
  \Gamma_S \lesssim \frac{1}{3 \times 2^8 \pi^3 (0.017)^2}
 \frac{m_S^3}{M_P T_{BL}}  ~,
\end{align}
and thus
\begin{equation}
 \tau_S \gtrsim 4 \times 10^{-3} \times\left(\frac{500~{\rm
				     GeV}}{m_S}\right)^3 
\left(\frac{T_{BL}}{10^{10}~{\rm GeV}}\right) ~{\rm s}~.
\end{equation}
Therefore, the lifetime of $S$ can be short enough to evade the BBN
bound. This result is qualitatively straightforward: a larger $S$ decay
rate requires larger couplings $y_{i}$ and $\lambda_i$, which cause
later decoupling of the wash-out processes. The $B-L$ generation should
then occur at a later time, which requires a low $T_{BL}$.

Finally we fit these particles into an SO(10) unification model
with an intermediate
gauge group as described in Sec.~\ref{SO(10)models}.  
In this specific model, the SO(10) gauge symmetry is broken by the VEV of a
singlet in ${\bf 210}_R$ into the intermediate symmetry group
$G_\text{int}=\text{SU}(4)_C \otimes \text{SU}(2)_L
\otimes\text{SU}(2)_R$ at $M_{\rm GUT}$. $G_\text{int}$ is then broken
down to the SM gauge group $G_\text{SM}$ at $M_{\rm int}$ by the VEV of
the $({\bf 10}, {\bf 1}, {\bf 3})_C$ component in a ${\bf 126}_C$ and
the $({\bf 15}, {\bf 1}, {\bf 1})_R$ component in the ${\bf 210}_R$. The
subscript $R$ and $C$ stand for real and complex fields,
respectively. The numbers in the parenthesis are the $G_\text{int}$
quantum numbers of the fields. The particle content except the SM
fermions and right-handed neutrinos are summarized in
Table~\ref{tab:extcogen}. There, the first column shows the particle
content around the electroweak or TeV scale. The second and
third columns show the quantum numbers under $G_\text{int}$ and the SO(10)
representation, respectively. $W$ stands for a Weyl field. Most of the
particles in the representation shown in the second column have
intermediate-scale masses, except for the components listed in the first
column. Other components of the SO(10) representation are assumed to lie
around $M_\text{GUT}$. The dark sector particles $\psi_S$, $\psi_D$ and
$S$ belong to Weyl $\bf 1$, Weyl $\bf 10$ and complex scalar $\bf 16$
representations, respectively. The SM Higgs doublet, which breaks
$G_\text{SM}$, is a mixture of doublets in $({\bf 1}, {\bf 2}, {\bf
2})_C$ of ${\bf 10}_C$ and $({\bf 10}, {\bf 2}, {\bf 2})_C$ of ${\bf
210}_C$. The latter component cannot couple to the SM fermions or
right-handed neutrinos but it can couple to the components of the ${\bf
16}_S$. At the intermediate scale $({\bf 10}, {\bf 2}, {\bf 2})_C$ and
$({\bf 15}, {\bf 1}, {\bf 1})_R$ are required only for achieving
unification. With this particle content, the one loop result of the
scales and unification coupling are  
\begin{equation}
M_\text{int}=10^{11.8}~{\rm GeV},\qquad M_\text{GUT}=10^{15.7}~{\rm GeV},
\qquad \alpha_\text{GUT}=0.027~.
\end{equation}
The high intermediate scale guarantees small neutrino masses by the
type-I seesaw mechanism. The constraint from proton decay is evaded by
the high unification scale.

\begin{table}[ht]
 \begin{center}
\caption{\it Particle content of the extended cogenesis model. The first
  column shows the particle content around the electroweak or TeV
  scale. The second column and the third column show the quantum number
  under $G_\text{int}$ and the SO(10) representation, respectively. }
\label{tab:extcogen}
\vspace{5pt}
\begin{tabular}{ccc}
\hline
\hline
~~~EW~~~~ & $\text{SU}(4)_C \otimes \text{SU}(2)_L \otimes
  \text{SU}(2)_R$~~ & SO(10) \\
\hline
 $S$ & $({\bf 4}, {\bf 1}, {\bf 2})_C$  & ${\bf 16}_C$ \\
 $\psi_D$ & $({\bf 1}, {\bf 2}, {\bf 2})_W$ & ${\bf 10}_W$ \\
 $\psi_S$ & $ ({\bf 1}, {\bf 1}, {\bf 1})_W$  & ${\bf 1}_W$ \\
  $H$    &  $({\bf 10}, {\bf 2}, {\bf 2})_C $    &  ${\bf 210}_R$ \\
  $H$    &  $({\bf 1}, {\bf 2}, {\bf 2})_C $    &  ${\bf 10}_C$ \\
         &  $({\bf 15}, {\bf 1}, {\bf 1})_R $    &  ${\bf 210}_R$ \\
          &  $({\bf 10}, {\bf 1}, {\bf 3})_R $    &  ${\bf 126}_C$ \\        
 \hline
\hline
\end{tabular}
 \end{center}
\end{table}

\subsection{Asymmetry transfer by Yukawa coupling}
\label{sec:stop-bino}

In this section we present another possibility where the asymmetry in
$X_2$ is obtained from a Yukawa coupling of the form
$X_2\overline{X}_1f$ with $f$ representing the SM fermions. $X_1$ and 
$X_2$ are taken to be a Majorana fermion and a complex scalar multiplet,
respectively. The chemical potentials of $X_1$ and $X_2$ are determined
by the neutrality of the Majorana particle $X_1$ and by this Yukawa
interaction, respectively:  
\begin{equation}
\mu_{X_1}=0,\qquad \mu_{X_2}=-\mu_f~.
\end{equation}

Similarly to the extended cogenesis model, $X_2$ is supposed to decay
into $X_1$ after the $X_1$--$X_1$ and $X_2$--$X_2^*$ annihilation
processes decouple. If these annihilation processes deplete $X_1$
and the symmetric part of $X_2$ density efficiently, the relic abundance
will be determined by the asymmetric part of the $X_2$ density before
its decay. As we will see, the slow decay $X_2\rightarrow X_1+\bar{f}$
requires a small mass gap between $X_2$ and $X_1$. 
At low temperature $T<m_{X_{1}}$, the asymmetry is transferred through
the scattering $f + X_{1,2}\rightarrow A+X_{2,1}$ with SM fermions
propagating in the $t$-channel and $A$ is any light gauge boson which couples to $f$. 
The decoupling temperature of the
asymmetric transfer $T_{\rm dec}$ is thus determined by the decoupling
of this $t$-channel scattering process.  

As a concrete example of this model, we choose $X_2$ as a right-handed
stop-like particle $\widetilde{t}_R$, which is a color triplet, weak
isospin singlet and has hypercharge $2/3$. $X_1$ is chosen as a
mixture of a singlet Majorana fermion $\psi_S$ and the neutral component
of a doublet $\psi_D^0$, as in the extended cogenesis model
discussed in the previous subsection. Furthermore, we assume
$\widetilde{t}_R$ only couples to the right-handed top quark $t_R$
through the Yukawa coupling
\begin{equation}
\mathcal{L} = \lambda_t\overline{t}_R\psi_S \widetilde{t}_R + {\rm
 h.c.}~,
\label{eq:lamt}
\end{equation} 
which resembles the bino-stop-top coupling in the minimal
supersymmetric Standard Model.

The SO(10) completion of this model on top of the three generations of the
SM $\bf 16$ is summarized in Table~\ref{tab:stoptrans}. The SO(10)
symmetry is broken by a ${\bf 210}_R$ to $G_\text{int}=\text{SU}(4)_C
\otimes \text{SU}(2)_L \otimes \text{SU}(2)_R$,  which is broken
subsequently to $G_{\rm SM}$ by the VEV of $({\bf 10}, {\bf 1}, {\bf
3})_C$ in a ${\bf 126}_C$. $\psi_S$, $\psi_D$ and $\widetilde{t}_R$
belong to Weyl $\bf 45$, Weyl $\bf 10$, and complex scalar $\bf 16$
representations, respectively.\footnote{We are required here to consider
a higher representation for $\psi_S$ to achieve gauge coupling
unification with a sufficiently high GUT scale. } 
The Yukawa interaction \eqref{eq:lamt}
comes from the coupling ${\bf 16}^*{\bf 16}_f {\bf 45}_W$ where ${\bf
16}_f$ is the multiplet composed of the third generation SM fermions and
right-handed neutrino. $G_\text{SM}$ is broken by the VEV of the
following doublets: $({\bf 1}, {\bf 2}, {\bf 2})_R$ of ${\bf 10}_R$,
$({\bf 15}, {\bf 2}, {\bf 2})_C$ of ${\bf 126}_C$, and $({\bf 10}, {\bf
2}, {\bf 2})_C$ of ${\bf 210}_R$. The SM Higgs doublet is a mixture of
the above doublets. The latter two multiplets at the intermediate scale
are necessary for achieving a sufficiently high unification scale. With this
particle content, the one-loop result for the scales and unification
coupling are  
\begin{equation}
M_\text{int}=10^{11.3}~{\rm GeV},\qquad M_\text{GUT}=10^{15.7}~{\rm GeV},
\qquad \alpha_\text{GUT}=0.035~.
\end{equation}

\begin{table}[ht]
 \begin{center}
\caption{\it Particle content of the stop mediated asymmetry transfer model. The first column shows the particle content around the electroweak or TeV scale. The second column and the third column show the quantum number under $G_\text{int}$ and the SO(10) representation respectively. }
\label{tab:stoptrans}
\vspace{5pt}
\begin{tabular}{ccc}
\hline
\hline
~~~EW~~~~ & $\text{SU}(4)_C \otimes \text{SU}(2)_L \otimes
  \text{SU}(2)_R$~~ & SO(10) \\
\hline
 $\widetilde{t}_R$ & $({\bf 4}, {\bf 1}, {\bf 2})_C$  & ${\bf 16}_C$ \\
 $\psi_D$ & $({\bf 1}, {\bf 2}, {\bf 2})_W$ & ${\bf 10}_W$ \\
 $\psi_S$ & $ ({\bf 1}, {\bf 1}, {\bf 3})_W$  & ${\bf 45}_W$ \\
  $H$    &  $({\bf 15}, {\bf 2}, {\bf 2})_C $    &  ${\bf 126}_C$ \\
  $H$    &  $({\bf 10}, {\bf 2}, {\bf 2})_C $    &  ${\bf 210}_R$ \\ 
  $H$    &  $({\bf 1}, {\bf 2}, {\bf 2})_C $     &  ${\bf 10}_C$ \\ 
        &  $({\bf 10}, {\bf 1}, {\bf 3})_C $     &  ${\bf 126}_C$ \\   
 \hline
\hline
\end{tabular}
 \end{center}
\end{table}

\begin{figure}[t]
\begin{center}
\includegraphics[width=0.6\textwidth]{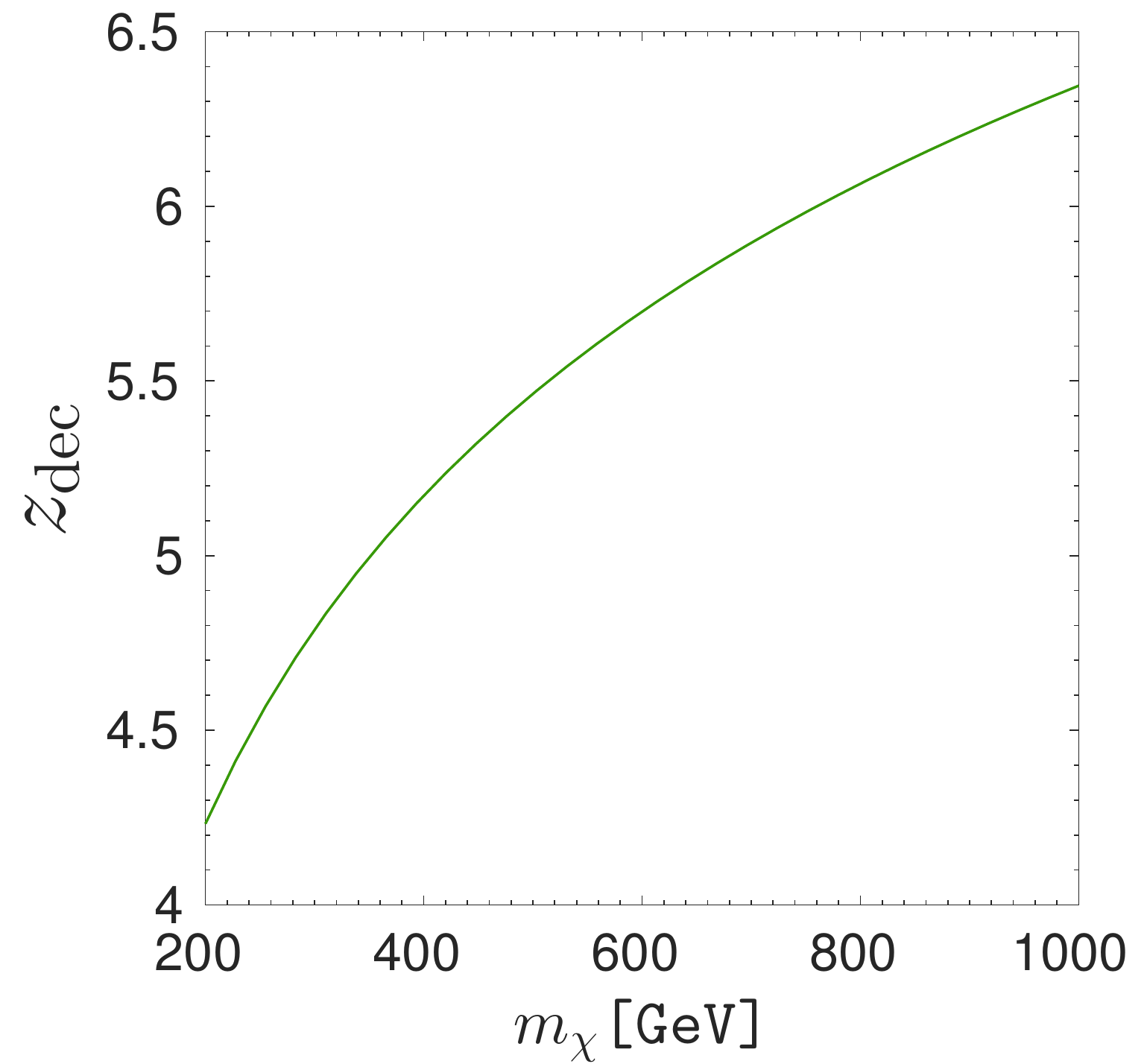}
\end{center}
\caption{{\it $z_{\rm dec}=m_{\tilde{t}_R}/T_{\rm dec}$ as a function of $m_{\chi}$, determined by dark matter relic density.}}
\label{fig:za_squark_bino}
\end{figure}

Now we consider the constraint placed on the coupling strength
$\lambda_t$ and the particle masses. The constraint on the mass of  $\chi$ from
efficient annihilation is the same as that discussed in
Sec.~\ref{sec:ExtCogenesis}.\footnote{As we see below, the coupling
$\lambda_t$ is required to be very small, and thus the contribution of
the interaction \eqref{eq:lamt} to the annihilation of the dark matter
particles is negligible. Moreover, since the conversion process $t\chi
\leftrightarrow g\widetilde{t}_R$ decouples before the decoupling of the
dark matter (see the discussion below), coannihilation with
$\widetilde{t}_R$ is ineffective. 
} The relationship between the relic density
and the dark matter mass is again given by Eq.~\eqref{eq:mdmformula},
with the relevant quantities for dark matter in
Eq.~\eqref{eq:mdmformula} replaced with the corresponding quantities for
$\widetilde{t}_R$; namely, we set $n_H =1$, $n_{\rm DM} =1$, $Y_{\rm DM} = 2/3$,
$X_{\rm DM} = 1/3$, and replace $k (z_{\rm dec})$ with $3k (z_{\rm
dec})$ to take the color factor for $\widetilde{t}_R$ into
account.\footnote{Strictly speaking, we may not directly apply
Eq.~\eqref{eq:mdmformula} to the present case as $\widetilde{t}_R$ can
be in thermal bath until the time of the sphaleron decoupling, though
this effect does not affect our discussion significantly. }
We then have
\begin{equation}
m_\chi \simeq m_N \left(\frac{\Omega_c}{\Omega_B}\right)
\frac{474+144k(z_{\rm dec})}{185k(z_{\rm dec})}~,
\end{equation}
where $z_{\rm dec}=m_{\tilde{t}_R}/T_{\rm dec}$ with $T_{\rm dec}$ the
decoupling temperature of the Yukawa interaction, and we have used Eq.~\eqref{eq:bsmbl}. The required value for
$z_{\rm dec}$ is then obtained from the observed dark matter density
using this relation, as shown in Fig.~\ref{fig:za_squark_bino}. We find
that it is in the range of 4--6.5 for $m_{\chi }$ of
$200~{\rm GeV}$--1~TeV. 

According to Fig.~\ref{fig:za_squark_bino}, around the decoupling
temperature of the Yukawa interaction \eqref{eq:lamt}, the temperature
is as low as $\sim 100$~GeV and thus even the dominant $t$-channel
scattering process $t+ \chi \leftrightarrow g + \widetilde{t}_R$, with $g$ a
gluon, is exponentially suppressed. The reaction rate is estimated as 
\begin{equation}
\Gamma (t \chi \leftrightarrow g  \widetilde{t}_R )\simeq
 \frac{g_3^2\lambda_t^2}{\pi m_\chi m_t}\cdot
\left(\frac{m_t T}{2\pi}\right)^{3/2} e^{-m_t/T}~.
\end{equation}
Through this process, any asymmetry in tops (baryon asymmetry) is transferred to an
asymmetry in the $\widetilde{t}_R$ which subsequently decay to $\chi$.
The decoupling temperature is estimated from $\Gamma (t
\chi\leftrightarrow g  \widetilde{t}_R )\simeq  H$ and using the result in
Fig.~\ref{fig:za_squark_bino}, we then obtain $\lambda_t \simeq
1.4~(1.1)\times 10^{-6}$ for $m_{\chi}=200~(1000){~\rm GeV}$.

Now let us consider the condition that $\widetilde{t}_R$ has a lifetime
long enough to decay after the annihilation of the symmetric part of
$\chi$ is over. To ensure such slow decay, we need to set  $\Delta m
\equiv m_{\widetilde {t}_R}-m_{\chi}<m_t$
so that the two-body decay channel
$\widetilde{t}_R\rightarrow t \chi$ is kinematically forbidden.\footnote{Such a small mass difference
also allows $\widetilde{t}_R$ to evade the strong limits from stop
searches at the LHC \cite{ATLAS:2016ljb, CMS:2016vew}.} The dominant
decay channel is then the three-body decay 
$\widetilde{t}_R\rightarrow bW\chi_i$~, $i=1,2,3$ represents three mass eigenstates
of $\psi_D^0-\psi_S$ mixing, and for simplicity we assume
$\widetilde{t}_R$ can decay to all of them, so that the decay rate is
not suppressed by the mixing angle. 
The decay occurs after $\chi$--$\chi$ annihilation if $\Gamma_{\tilde{t}_R}<H|_{T_f}$, where 
$m_\chi/T_f\sim20$ is the decoupling temperature of the annihilation.
Numerical calculation of the decay rate gives a bound of 
$\Delta m\lesssim 110~(160)~{\rm GeV}$ for
$m_{X_1}=200~(1000)~{\rm GeV}$, assuming the three $\chi_i$ are degenerate in mass.

Since $\widetilde{t}_R$ has a relatively long lifetime,\footnote{The lifetime for the mass ranges considered is
$\mathcal{O}(1)$ ns, and so is clearly not a problem for BBN.} it is hadronized before it
decays when produced at colliders. Such a massive charged particle
({\it e.g.}, $R$-hadron) yields characteristic signatures in the
detectors. The ATLAS \cite{Aaboud:2016uth} and CMS
\cite{Khachatryan:2016sfv} collaborations have searched for heavy
charged stable particles based on observables related to large ionization energy losses. Both searches require the $R$-hadron to get to the calorimeter in order to pass data selection, and this corresponds to a lifetime $\sim 4{~\rm ns}$.
The constraint for charged stable $R$-hadrons sets a lower bound on
stable stop and sbottom masses up to $\sim 900{~\rm GeV}$. This implies
an upper limit to the $\widetilde{t}_R$ lifetime,
$\tau_{\widetilde{t}_R}\lesssim 4{~\rm ns}$, otherwise the density originating from an asymmetry
can only occupy up to $20\%$ of the relic density
when $m_{\tilde{t}_R}\sim 900{~\rm GeV}$. A more severe limit on
$\widetilde{t}_R$ may in principle be obtained from the displaced vertex
searches \cite{Aad:2015uaa}. However, we cannot directly apply the
results in Ref.~\cite{Aad:2015uaa} to the present case due to the small
mass difference $\Delta m$. Although the reconstruction efficiency of
displaced vertices remains sizable even if $\Delta m \lesssim 100$~GeV
\cite{Nagata:2015hha}, the trigger efficiency is reduced, which results
in a weaker constraint. A dedicated study of this limit is beyond the
scope of the present paper.

Finally we remark that the framework of transferring the asymmetry through Yukawa
interactions can also be applied straightforwardly to other choices of
$\mathbb{Z}_2$-odd particles. For example, we can also choose $X_1$ as a single
Majorana triplet (thus avoiding the need for mixing among two multiplets) and 
$X_2$ as a slepton-like doublet. The asymmetry is
transferred to $X_2$ from the lepton doublet. The decoupling temperature
of asymmetry transfer in this case is however exponentially sensitive to
the Yukawa coupling, since the asymmetry transfer scattering
$X_1+\ell^-\leftrightarrow \gamma+X_2$ is mediated by a lepton and the
reaction rate is dependent on ${\rm log}(T/m_\ell)/T$ when $m_\ell\ll T\ll X_1
$. We will not discuss this model in more detail here.

\section{Conclusion}
\label{sec:conclusion}

The baryon and relic dark matter densities are known to be
quite similar. It is not known, however, whether or not the origin of 
these densities are related. 

The standard thermal mechanism for obtaining the correct dark matter
relic abundance is quite robust. The annihilation of dark matter
candidates with weak-scale (or TeV) masses with weak-scale (or slightly
weaker) interactions leaves behind a density close to that determined
observationally. Of course there are many non-thermal mechanisms which
may also produce the correct relic density. 
The calculation of the baryon density on the other hand requires
a model of baryogenesis and necessarily relies on unknown quantities such
as C and CP violating phases. Among the many mechanisms for generating the 
baryon asymmetry, one of the most attractive (and simplest) is the
out-of-equilibrium decay of a heavy right-handed neutrino as in the
original leptogenesis scenario \cite{fy}. While this scenario also relies
on the values of unknown phases, very little is needed beyond the
existence of the heavy right-handed states already included in SO(10)
models and responsible for the observed low-mass neutrinos. If the
generation of the baryon (or lepton) asymmetry is accompanied by the
simultaneous generation of an asymmetry in dark matter \cite{ho}
(assuming then that the symmetric component is driven to a low value
through thermal annihilation), we can understand why these two numbers
are close.

In this paper, we have attempted to construct a model of asymmetric
SO(10) dark matter. In addition to being able to account for neutrino
masses and leptogenesis quite naturally, SO(10) models which break
through an intermediate scale gauge group contain an unbroken
$\mathbb{Z}_2$ symmetry which can account for the stability of dark
matter. This is similar to $R$-parity in supersymmetric models, but
appears directly from  gauge symmetry breaking. Utilizing the presence
of an intermediate scale, these SO(10) models
may account for gauge coupling unification without supersymmetry. 
However despite the many possible constructions of models with
different intermediate scale gauge groups and dark matter representations, 
there are relatively few models which allow gauge coupling unification
with phenomenologically acceptable intermediate and GUT mass scales \cite{noz}. 
The question we posed here, is whether any model can be constructed for which
the dark matter density is connected to the baryon asymmetry.

From the list of possible SO(10) dark matter candidates, only a subset of them
can accommodate an asymmetry (for example all Majorana fermion
candidates are excluded). Furthermore, the universe of candidates is
further limited when constraints from direct detection experiments are
applied. We must also require that the symmetric thermal component
annihilate efficiently. We first considered minimal models, where the
field content is limited to the  SM matter representations, the Higgs
representations needed to break the GUT, intermediate and SM gauge
symmetries, along with a single dark matter representation. For minimal models,
we argued that the only candidate for asymmetric dark matter is the
complex scalar singlet residing in a {\bf 16} or a {\bf 144} of SO(10). 
We found that the constraints from annihilation, and direct detection precluded 
the ability to transfer the asymmetry from the SM sector to the dark matter
sector through thermal interactions unless the intermediate scale is
relatively low (see Eq.~\eqref{lowint}). However, without further
complicating the model, gauge coupling unification would be lost. In
contrast, the addition of a single fermion singlet would allow the
cogenesis mechanism to simultaneously generate the dark matter
asymmetries along with the lepton asymmetry produced during
leptogenesis.

We also considered two extended models in which the asymmetry
in some field (labelled $X_2$ here) is generated from (in the case of the 
transfer mechanism) or with (in the case of cogenesis) the baryon and
lepton asymmetry. In the cogenesis model, the asymmetry in the scalar
singlet (from the {\bf 16}) decays to the dark matter which is a mixed state of
the SO(10) singlet and a weak bi-doublet from a {\bf 10} of SO(10).
The intermediate and GUT scales in this model are sufficiently high
to easily produce light neutrino masses through the see-saw and provide sufficiently long proton
lifetimes. In the specific transfer model described, the baryon asymmetry (stored in top quarks)
is transferred to a right-handed scalar color triplet (also in a {\bf 16}) which decays to the dark 
matter which is again a mixed state of a SM singlet (though now in a {\bf 45} of SO(10))
and the same weak bi-doublet
from a {\bf 10}. This model also has sufficiently high intermediate and GUT scales.

The fact that present-day experimental constraints, particularly from direct detection
experiments place strong constraints on these models, there remains hope that
these experiments will shed further light on the nature of dark matter and whether or not
the dark matter may be ultimately related to the baryon density of the Universe.

\section*{Acknowledgments}

This work
was supported in part by DOE grant DE-SC0011842 at the University of Minnesota.

\end{document}